\documentclass[superscriptaddress,onecolumn,letterpaper,aps]{revtex4-1}
\usepackage{amsmath}
\usepackage{amssymb}
\usepackage{graphicx}

\usepackage[T1]{fontenc}

\newcommand{\mb}[1]{\mathbf{#1}}
\newcommand{\bs}[1]{\boldsymbol{#1}}

\begin{document}

\title{Anisotropic Hydrodynamic Mean-Field Theory for Semiflexible
  Polymers under Tension}

\author{Michael Hinczewski}
\affiliation{Department of Physics, Technical University of  Munich, 85748 Garching, Germany}
\affiliation{Institute for Physical Science and Technology, University of Maryland, College Park, MD 20742}
\email{mhincz@umd.edu}
\author{Roland R. Netz}
\affiliation{Department of Physics, Technical University of Munich, 85748 Garching, Germany}
\affiliation{Fachbereich Physik, Arnimallee 14, Freie Universit\"at Berlin, 14195 Berlin, Germany}

\hyphenation{non-e-qui-lib-ri-um}

\begin{abstract}
We introduce an anisotropic mean-field approach for the dynamics of
semiflexible polymers under intermediate tension, the force range
where a chain is partially extended but not in the asymptotic regime
of a nearly straight contour.  The theory is designed to exactly
reproduce the lowest order equilibrium averages of a stretched
polymer, and treats the full complexity of the problem: the resulting
dynamics include the coupled effects of long-range hydrodynamic
interactions, backbone stiffness, and large-scale polymer contour
fluctuations.  Validated by Brownian hydrodynamics simulations and
comparison to optical tweezer measurements on stretched DNA, the
theory is highly accurate in the intermediate tension regime over a
broad dynamical range, without the need for additional dynamic fitting
parameters.
\end{abstract}

\maketitle

\section{Introduction}

Understanding semiflexible biopolymers under tension is relevant both
{\it in vivo}---determining the mechanical response of cytoskeletal
networks \cite{Mizuno2007,Bendix2008,Schmoller2008}---and {\it in
  vitro}, due to a proliferation of single-molecule techniques like
optical tweezers that involve manipulating stretched
DNA~\cite{Meiners2000,Woodside2006,Gebhardt2010}.  Among the basic
goals of current research is to achieve a comprehensive,
quantitatively accurate theory for the dynamical response of a
semiflexible chain prestretched by an external force.  For polymers
under certain asymptotic conditions this effort has been highly
successful: where the persistence length $l_p$ is much greater than
the total length $L$, or for a large force $F \gg k_BT /l_p$, we can
assume the polymer contour remains nearly straight, and apply the
weakly bending approximation (WBA)~\cite{Granek1997}.  This has proven
a versatile approach, useful in both
equilibrium~\cite{Granek1997,Caspi1998,Hiraiwa2008} and
non-equilibrium~\cite{Bohbot-Raviv2004,Hallatschek2005,Obermayer2007,Obermayer2009}
contexts for stretched semiflexible polymers.  In the context of
single-molecule experiments, the WBA can work very well for actin
filaments~\cite{LeGoff2002}, where the regime $L \lesssim l_p \sim
{\cal O}(10\:\mu\text{m})$ is easily accessible.

However, to complete the dynamical picture, we need a complementary
approach for cases that do not fall within the weakly bending regime.
Away from the asymptotic limit, when dealing with weaker forces or
more flexible chains, we are confronted by complex crossovers between
dynamical regimes at short times (dominated by backbone rigidity) and
those at larger time scales, where flexible chain modes come into
play.  Adding to the complexity is the role of long-range hydrodynamic
interactions between polymer segments.  For weakly bending chains,
these can be approximately incorporated by assuming distinct
longitudinal and transverse friction coefficients, $\zeta_\parallel
\approx \zeta_\perp/2$, appropriate for a rigid
rod~\cite{Hallatschek2005,Nyrkova2007,Obermayer2009}.  This
renormalizes times scales, without affecting the dynamic scaling.
(While the assumption of distinct friction coefficients remains the
most common approximation, one can also use a more complicated
pre-averaging approach~\cite{Winkler2007}.)  When the chain is not
weakly bending, simple rod-like hydrodynamics is no longer valid, and
we need another method of dealing with the long-range coupling.

The experimental significance of the non-asymptotic regime has been
highlighted by optical tweezer single-molecule applications involving
small stretching forces, $F \sim {\cal O}(10^{-1} - 10)$ pN, and more
flexible polymers like double-stranded DNA, where $l_p \approx 50$ nm
and the typical strand lengths $L \sim {\cal O}(10^2 -
10^4)\:\text{nm} \gtrsim
l_p$~\cite{Meiners2000,Woodside2006,Gebhardt2010}.  In order to
quantitatively capture such experimental setups, we need a theory that
bridges the flexible, zero-force regime of classical polymer
approaches like the Zimm model~\cite{Zimm1956,DoiEdwards}, and the
strongly-stretched, stiff limit where the WBA is successful (which in
the DNA case when $L\gtrsim l_p$ requires $F \gg k_B T/l_p \approx
0.08$ pN).

The current work focuses on addressing this need, by constructing an
anisotropic mean-field theory (MFT) for a semiflexible chain under
tension, including hydrodynamics through a pre-averaging
approximation.  To verify the theory, we also carry out extensive
comparisons with bead-spring worm-like chain Brownian hydrodynamics
(BD) simulations.  The simulation results for flexible, partially
extended chains underscore the complexity of polymer dynamics in the
non-asymptotic case: quantities like the mean squared displacement
(MSD) of the chain end-point or end-to-end distance show broad
crossovers between short and long-time dynamics, rather than distinct
regimes characterized by simple power law scaling.  Moreover, by
comparing numerical results with and without long-range hydrodynamics,
we find that long-range coupling through the solvent does have a
significant effect---one that must be included in any theoretical
approach to obtain a quantitative comparison with experiments.

Remarkably, the anisotropic mean-field theory captures
both the crossover and hydrodynamic effects, giving excellent
agreement with the simulations.  In fact, the solvent-mediated
coupling between all points on the polymer is not an incidental
element in the method, but a key to its success: the long-range
interactions make the mean-field approach more realistic.  This has
already been demonstrated for $F=0$, where an earlier, isotropic
MFT~\cite{Hinczewski2009,Hinczewski2009b} was able to model
precisely the dynamics of an end-monomer in DNA strands observed
through fluorescence correlation spectroscopy (FCS)~\cite{Petrov2006};
the experimental validation is reviewed in Sec.~\ref{iso} below.  A
salient aspect of the experimental comparison was that the isotropic MFT
required no fitting parameters: starting from the known properties of
the system, it could reproduce the measured end-monomer MSD over five
decades of time and three decades of chain lengths.  While we do not
yet have comparably detailed experimental results for single chains
under tension, we are able to check our anisotropic MFT against
measurements of longitudinal and transverse relaxation times of
partially extended DNA molecules in an optical
tweezer~\cite{Meiners2000}.  Again we can match the experimental
results without dynamic fitting parameters, which is a non-trivial test of our
approach, since the relaxation times are sensitive to details like the
hydrodynamic coupling in the chains.

The reason for resorting to an anisotropic, rather than an isotropic,
mean-field approach at $F\ne 0$ arises from limitations revealed in
earlier attempts to incorporate prestretching tension using an
isotropic Hamiltonian~\cite{Ha1997,Harnau1999b,Winkler2003}.  An
isotropic theory cannot reproduce the distinct equilibrium
thermodynamic averages for directions parallel and perpendicular to
the applied force.  A prerequisite for a good dynamical theory is that
it must yield the correct equilibrium properties in the long time
limit.  With this in mind, the Hamiltonian of our anisotropic theory
is designed to give the {\it exact} lowest-order equilibrium averages
for a stretched semiflexible chain---derived from the numerical
quantum solution of the worm-like chain (WLC) model.  After fixing the
correct static quantities, we use our theory to predict dynamical
quantities: amplitudes and relaxation times of the chain fluctuation
modes, and related physical observables like the end-point and
end-to-end MSDs.  Among the interesting qualitative viscoelastic
properties we find is semiflexible polymer stiffening under tension,
which has been seen experimentally in cytoskeletal networks put under
stress either through deformation~\cite{Caspi1998}, or the activity of
motor proteins~\cite{Mizuno2007}.

The anisotropic MFT is complementary to the WBA, in the sense that it
is most accurate in regimes where the WBA breaks down.  Conversely,
certain aspects of the mean-field approximation---like the predicted
longitudinal dynamics---do not work in the asymptotic weakly bending
limit.  (In contrast the MFT transverse dynamics reduce to the
conventional WBA results in this limit, up to hydrodynamic corrections
which are included in our MFT treatment.)  This is not surprising,
since a Gaussian model, like the one arising from our mean-field
approach, can never capture the longitudinal fluctuations of a nearly
rigid rod.  However, while there are many good theories for the
asymptotic regime, the non-asymptotic case is less well understood,
and this is where our method will be most useful.  Given the
resolution of current single-molecule experimental techniques, one
should be able to sensitively probe the fine details of dynamical
behavior predicted by our theory and simulations, including crossover
and hydrodynamic effects over a broad range of time scales.

The paper is organized as follows: Sec.~\ref{iso} summarizes the
earlier development of the isotropic MFT.  For $F=0$, we focus on the
predictive power of the theory for single-molecule experiments on DNA
dynamics.  However, generalizing this success to $F \ne 0$ turns out
to be fraught with difficulties.  In Sec.~\ref{amft} we present a
resolution to the problem, introducing an anisotropic MFT Hamiltonian
and deriving the corresponding dynamical theory, based on a
hydrodynamic pre-averaging approach.  Sec.~\ref{bd} describes the BD
simulations used to check the theory.  Finally, Sec.~\ref{results}
presents results: comparisons with simulations
(Sec.~\ref{results:bdcomp}), with an optical tweezer experiment on DNA
(Sec.~\ref{results:dnacomp}), and with the WBA
(Sec.~\ref{results:wbacomp}).
 
\section{Strengths and limitations of the isotropic MFT}\label{iso}

We begin by reviewing the isotropic MFT approach to semiflexible
polymers at
$F=0$~\cite{Winkler1994,Ha1995,Harnau1996,Hinczewski2009,Hinczewski2009b}
and $F\ne 0$~\cite{Ha1997,Harnau1999b,Winkler2003}.  The starting
point is the WLC Hamiltonian,
\begin{equation}\label{iso1}
U_\text{WLC} = \frac{l_pk_B T}{2} \int ds\, (\partial_s \mb{u}(s))^2 - F
\hat{\mb{z}}\cdot \int ds\,\mb{u}(s),
\end{equation}
which describes the elastic energy of a space curve $\mb{r}(s)$ with
contour coordinate $0 \le s \le L$ and tangent vector $\mb{u}(s)
\equiv \partial_s \mb{r}(s)$ constrained by local inextensibility to
$|\mb{u}(s)|=1\: \forall s$.  The first term is the bending energy,
parametrized by the persistence length $l_p$, while the second term is
the external field due to a force $F$ along the $z$ axis.  The
$|\mb{u}(s)|=1$ constraint makes the dynamics of the system
analytically intractable, but the partition function $Z$, expressed as
a path integral over $\mb{u}(s)$, can be approximated through the
stationary phase approach, yielding a Gaussian mean-field model.  The
end result is
\begin{equation}\label{iso2}
Z_\text{MF} = \exp(-\beta {\cal F}_\text{MF}) = \int {\cal
  D}\mb{u}\, \exp(-\beta U_\text{MF}),
\end{equation}
 where local inextensibility has been relaxed and the MFT Hamiltonian
 is:
\begin{equation}\label{iso3}
U_\text{MF}=\frac{\epsilon}{2} \int ds\,\left(\partial_s
 \mb{u}(s)\right)^2 +\nu \int ds\,\mb{u}^2(s) + \nu_0
 \left(\mb{u}^2(0) + \mb{u}^2(L)\right) - F \hat{\mb{z}} \cdot \int
 ds\,\mb{u}(s).
\end{equation}
Here $\epsilon \propto l_p k_BT$ parametrizes the bending energy, and
the new terms parametrized by $\nu$ and $\nu_0$ are bulk and end-point
stretching energies respectively.  From the stationary phase condition,
$\partial_{\nu} {\cal F}_\text{MF} = \partial_{\nu_0} {\cal
  F}_\text{MF} = 0$, one finds that the parameters $\nu$ and $\nu_0$
are functions of $F$ and $\epsilon$, and act as Lagrange multipliers
enforcing the global and end-point constraints $\int ds\, \langle
\mb{u}^2(s)\rangle = L$, $\langle \mb{u}^2(0) \rangle = \langle
\mb{u}^2(L)\rangle = 1$.
\subsection{Isotropic MFT at $F=0$}

\begin{figure}[!t]
\centering\includegraphics*[scale=1]{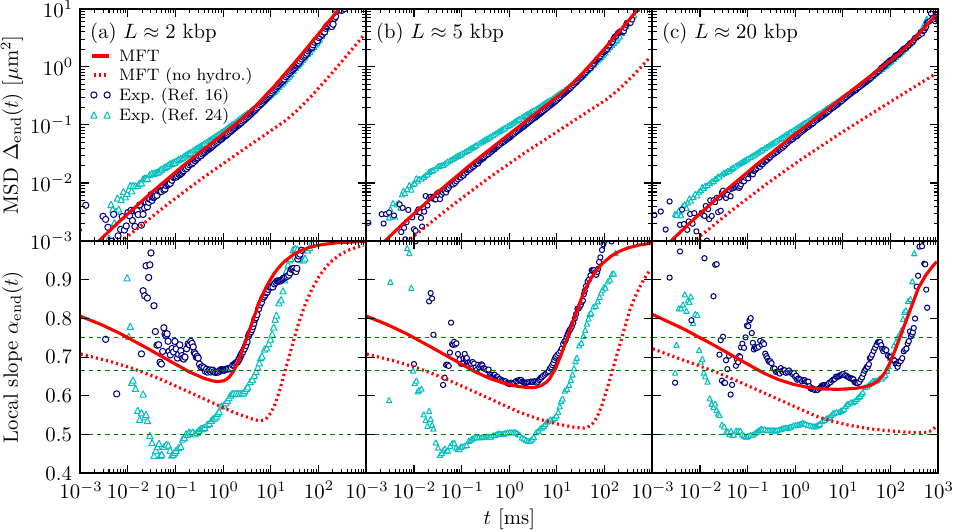}
\caption{Top panels: The mean squared displacement (MSD),
  $\Delta_\text{end}(t)$, of a fluorescently tagged end-point in a
  double-stranded DNA chain of contour length (a) $L \approx 2$ kbp,
  (b) $L \approx 5$ kbp, (c) $L \approx 20$ kbp.  Bottom panels: the
  local slope $\alpha_\text{end}(t) = d\ln \Delta_\text{end}(t)/d\ln
  t$, calculated by linear fitting to the log-log plots of
  $\Delta_\text{end}(t)$ within a time window $t_i$ around each $t$
  defined by $|\log_{10} t_i/t| < 0.15$.  The symbols show the results
  of two different FCS measurements: Shusterman {\it
    et.~al.}~\cite{Shusterman2004} (triangles) and Petrov {\it
    et.~al.}~\cite{Petrov2006} (circles).  Comparable values of $L$
  are chosen from each experiment: 2400, 6700, 23100 bp from
  Ref.~\citenum{Shusterman2004} and 1965, 5058, 19941 bp from
  Ref.~\citenum{Petrov2006}.  The isotropic hydrodynamic MFT
  results~\cite{Hinczewski2009b,Hinczewski2010}, without fitting
  parameters, are drawn as solid red curves (with $L$ values matching
  the Ref.~\citenum{Petrov2006} data).  For comparison, to highlight
  the importance of long-range hydrodynamic interactions, the results
  of the free-draining MFT theory without hydrodynamics
  are drawn as dotted red curves.  The small deviations in $L$ between
  experiments do not lead to significant changes in the MFT curves on
  the scale of the figure---thus the hydrodynamic MFT clearly agrees
  with Petrov {\it et.~al.}  rather than Shusterman {\it et.~al.}.
  The horizontal dashed lines in the bottom panel show power-law
  exponent values from various scaling theories: the Rouse model
  ($\alpha_\text{end} = 1/2$), the Zimm model
  ($\alpha_\text{end}=2/3$), and the worm-like chain
  ($\alpha_\text{end}=3/4$).}
\label{exper}
\end{figure}

For $F=0$, $U_\text{MF}$ has been studied
extensively~\cite{Winkler1994,Ha1995}, and setting $\epsilon =
(3/2)l_p k_BT$ it reproduces exactly the WLC tangent-tangent
correlation $\langle \mb{u}(s) \cdot \mb{u}(s^\prime) \rangle$ and
related quantities.  To extract dynamics, one can adopt a Zimm-like
hydrodynamic pre-averaging
approach~\cite{Harnau1996,Hinczewski2009,Hinczewski2009b}, described
in more detail below in the context of the anisotropic MFT.  Within
this approach, the chain contour obeys a Langevin equation that can be
solved through normal mode decomposition.  The power of the resulting
theory is amply illustrated through an experimental example, where the
accuracy of the predicted dynamics was able to resolve a
conflict between two FCS studies on
DNA~\cite{Hinczewski2009,Hinczewski2009b,Hinczewski2010}.

Both studies, by Shusterman {\it et.~al.}~\cite{Shusterman2004} and
later by Petrov {\it et.~al.}~\cite{Petrov2006}, measured the mean
squared displacement (MSD) of a fluorescent tag attached to the end of
double-stranded DNA molecules.  A variety of contour lengths $L = 0.1
- 20$ kbp ($\approx 30-7000$ nm) were used, three of which are shown
in \ref{exper} (taking comparable values of $L$ from each experiment).
In the top panels the end-point MSD, $\Delta_\text{end}(t)$, is
plotted as a function of time $t$ on a log-log scale, while the bottom
panels show the local slopes $\alpha_\text{end}(t) = d\log
\Delta_\text{end}(t)/d\log t$.  These slopes are useful in
characterizing the scaling behavior of the MSD, which is typically
analyzed in terms of power-law exponents.  A pure power-law scaling
would manifest itself as plateau with constant $\alpha_\text{end}(t)$,
but what we see instead is a continuous variation of the local slope,
a reflection of slow crossovers between different dynamical regimes.

Strikingly, the MSD curves from the two studies show a strong
divergence at small and intermediate times (at the largest time
scales, both converge to the same simple behavior, dominated by the
center-of-mass diffusion of the entire chain, with
$\alpha_\text{end}(t) \to 1$).  The Shusterman {\it et.~al.} data
exhibits an ``intermediate Rouse regime'', becoming more prominent for
longer chains, where $\alpha_\text{end}(t) \approx 1/2$ over times
where $l_p^2 \lesssim \Delta_\text{end}(t) \lesssim L^2$.  (The
persistence length $l_p \approx 50$ nm or 150 bp for DNA.)  This
surprising $t^{1/2}$ scaling of the MSD agrees with the Rouse model,
valid for flexible polymers in the absence of long-range hydrodynamic
coupling.  For the dilute solutions used in the experiments, where
screening by neighboring chains is negligible, the classical
expectation is that hydrodynamic corrections are significant.  Thus
one should see instead the $t^{2/3}$ scaling predicted by the Zimm
model~\cite{DoiEdwards}.  In fact, the Petrov {\it et.~al.} data does
show $\alpha(t)$ closer to the Zimm value at intermediate times, and
moreover the magnitude of the MSD is 2-3 times smaller than in the
first experiment.

The isotropic MFT results for $\Delta_\text{end}(t)$ and
$\alpha_\text{end}(t)$, including long-range hydrodynamic
interactions, are plotted as solid red curves in \ref{exper}.  There
are no fitting parameters, with all the constants taken either
directly from the experimental setup or the literature: $T = 298$ K,
$\eta = 0.891$ $\text{mPa}\cdot\text{s}$, $a = 1$ nm, a rise per bp of
$0.34$ nm, $l_p = 50$ nm.  The hydrodynamic MFT clearly distinguishes
between the two experiments, showing close agreement with the Petrov
{\it et.~al.}  results.  Using the full data set from the Petrov {\it
  et.~al.}  experiment (comparison shown in
Ref.~\citenum{Hinczewski2009b}) reveals that the hydrodynamic MFT provides a global
description of the DNA end-point dynamics, covering three decades of
strand length ($L \approx 30-7000$ nm), and five decades of time
($10^{-2} - 10^{3}$ ms).  Over the time scales where there is least
experimental uncertainty, $t = 10^{-1}-10^2$ ms, the average deviation
between theory and experiment ranges between $6-25\%$ for the
different $L$.

To achieve this level of accuracy without fitting parameters, the full
physical complexity of the problem must be taken into account,
particularly the off-diagonal coupling between normal modes due to
hydrodynamics.  The significance of hydrodynamic effects can be seen
by plotting the non-hydrodynamic isotropic MFT theory
for comparison (dotted red curves in \ref{exper}).  As expected, these
show $\alpha(t)$ values much closer to the Rouse prediction of 1/2,
with a clear asymptotic Rouse regime developing for longer $L$.
However the Shusterman {\it et.~al.} results do not match the
non-hydrodynamic theory either, with the experimental relaxation times
being smaller, and the observed MSD values actually higher than the
hydrodynamic ones at small times.  The opposite is true for the
theory: with long-range hydrodynamics screened, the MSD is noticeably
smaller than in the hydrodynamic case, since the effective solvent
friction felt by the chain is larger.  The lack of agreement between
the Shusterman {\it et.~al.}  data and either the hydrodynamic or
non-hydrodynamic theory points to an underlying issue with either the
setup or analysis involved in that study.

Overall, this example underlines the strengths of the MFT approach: it
can quantitatively reproduce the results of a single-molecule
experiment, down to non-trivial crossover behavior of the polymer at
different time scales.  This is a considerable success, given the
interplay of various effects reflected in the MSD curves of
\ref{exper}: (i) the backbone rigidity, dominant for
$\Delta_\text{end}(t) \lesssim l_p^2$, giving an exponent of 3/4 plus
hydrodynamic corrections, though not clearly resolved due to
uncertainties in the short-time experimental data; (ii) the Zimm-like
flexible intermediate regime, though with the crossover inducing an
exponent slightly smaller than 2/3 for longer
chains~\cite{Hinczewski2009}; (iii) the large-scale polymer motions at
long times, which include rotational and translational center-of-mass
diffusion.  All of these are reasonably described by the isotropic
MFT.

\subsection{Isotropic MFT at $F\ne 0$}

The motivation of the current study is to construct a theory that can
match the quantitative accuracy of the above example, but in the
presence of tension.  While a prestretching force can be simply
incorporated into the isotropic MFT, the results are mixed.  On the
one hand, the $F\ne 0$ isotropic MFT successfully yields the known
asymptotic forms for the average end-to-end extension parallel to the
force~\cite{Ha1997}:
\begin{equation}\label{iso4}
\begin{split}
\frac{\langle R_z \rangle}{L} &= \frac{2 l_p F}{3k_B T}, \qquad F \to 0,\\
\frac{\langle R_z \rangle}{L} &= 1 - \sqrt{\frac{3 k_B T}{ 8 l_p F}}, \qquad F \to \infty,
\end{split}
\end{equation}
where $\mb{R} = \int_0^Lds\, \mb{u}(s)$ is the end-to-end vector.
These agree with the Marko-Siggia exact result for the
WLC~\cite{Marko1995}, except for the factor $3/8$, which should be
1/4.

\begin{figure}[!t]
\includegraphics*[scale=1]{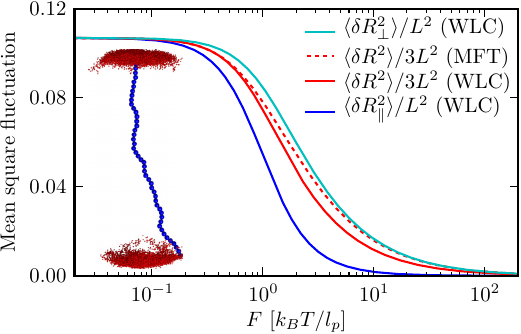}
\caption{For a polymer with $L/l_p = 5$, the end-to-end
  vector fluctuation $\langle \delta R^2 \rangle$ and its components,
  $\langle \delta R_\parallel^2 \rangle$, $\langle \delta R_\perp^2
  \rangle$, with varying $F$ (derived from the exact quantum solution
  of the WLC), compared to the isotropic MFT prediction.  Inset: cloud
  of end-point positions (red points) taken from a BD simulation of a
  polymer ($N=50$ beads), with $L/l_p = 5$ and force $F = 20\: k_B
  T/l_p$ applied along the vertical axis.  The initial polymer
  configuration is shown in blue.}\label{isomft}
\end{figure}

However, underlying this seemingly small discrepancy is a serious
problem in the isotropic MFT: for large $F$ it cannot correctly
account for the anisotropic fluctuation behavior of the WLC.  To
illustrate this, let us consider displacements $\delta R_\parallel =
R_z - \langle R_z \rangle$ and $\delta R_\perp = R_x$ or $R_y$.  The
MFT does not differentiate between $\parallel$ and $\perp$
fluctuations: $\langle \delta R_\parallel^2 \rangle = \langle \delta
R_\perp^2 \rangle$ for all $F$.  However for the WLC, $\langle \delta
R_\parallel^2 \rangle$ becomes much smaller than $\langle \delta
R_\perp^2 \rangle$, as is evident in the inset of \ref{isomft},
showing a snapshot of end-point fluctuations taken from a BD
simulation.  A similar anisotropy exists between the $\perp$ and
$\parallel$ fluctuations at $F=0$ for a stiff filament with $L
\lesssim l_p$ \cite{Everaers1999}.  Using the exact mapping of the WLC
onto quantum motion over the surface of a unit
sphere~\cite{Samuel2002}, we numerically calculate the fluctuation
magnitudes, which are shown in \ref{isomft}.  (For details see
Sec.~\ref{amft:quant}.)  Since the MFT averages over all coordinate
directions, its estimate for the magnitude converges to the exact
$\langle \delta R_\perp^2 \rangle$ at large $F$, as the $\perp$
fluctuations dominate in that regime.  It misses entirely the
$\parallel$ component.  Thus to understand the dynamic response of
stretched semiflexible polymers one needs a more suitable theoretical
starting point.

\section{Anisotropic MFT}\label{amft}

To resolve these difficulties, we propose an anisotropic version of the
Gaussian model:
\begin{equation}\label{amft1}
\begin{split}
&U_\text{MFA}=\sum_{\alpha = \parallel, \perp} \biggl[\frac{\epsilon_\alpha}{2} \int ds\,\left(\partial_s \mb{u}_\alpha(s)\right)^2 +\nu_\alpha \int
ds\,\mb{u}_\alpha^2(s) + \nu_{0\alpha} \left(\mb{u}_\alpha^2(0) + \mb{u}_\alpha^2(L)\right)\biggr] - \chi F \int ds\,u_\parallel(s)\,,
\end{split}
\end{equation}
with $u_\parallel = u_z$, $\mb{u}_\perp = (u_x,u_y)$.  The form of
$U_\text{MFA}$ follows from the isotropic mean-field Hamiltonian in
Eq.~\eqref{iso3} by breaking the symmetry between directions parallel
and perpendicular to the pulling force.  In addition to the six
parameters which arise from the bending $(\epsilon_\parallel,
\epsilon_\perp)$, bulk stretching $(\nu_\parallel,\nu_\perp)$, and
end-point stretching $(\nu_{0\parallel},\nu_{0\perp})$ terms, we have
the force term, which is renormalized by a factor $\chi$.  The guiding
philosophy will be similar to the isotropic case: a dynamical theory
based on a Hamiltonian closely approximating the equilibrium behavior
of the WLC under tension.  For this purpose, we require that
$U_\text{MFA}$ should reproduce exactly the following lowest-order WLC
averages: $\langle R_\parallel \rangle$, $\langle \delta R_\alpha^2
\rangle$, $\int_0^L ds\,\langle u_\alpha^2(s) \rangle$, $\langle
u_\alpha^2(0)+u_\alpha^2(L) \rangle$, $\alpha = \parallel, \perp$.
The latter can be calculated from the quantum solution to the WLC
(Sec.~\ref{amft:quant}).  Since there are a total of seven of these
averages, we need seven free parameters in $U$ so that the theory can
exactly match all the WLC results.  (Hence the presence of the force
rescaling factor $\chi$, which is not otherwise required by
symmetry-breaking.)  The analytical expressions for these averages
derived from $U$ lead to seven equations for the seven unknown
parameters.  These can be solved numerically for any given $L$, $l_p$,
and $F$ (examples are shown in Sec.~\ref{amft:par}).  In the limit $F
\to 0$, our approach recovers the stationary phase condition of the
$F=0$ isotropic model, as expected.

Our dynamical theory builds on the hydrodynamic pre-averaging approach
used earlier for the isotropic MFT~\cite{Harnau1996,Hinczewski2009}.
Here we give a brief outline of the approach, with the full details in
Sec.~\ref{amft:dyn}.  The time evolution of the chain $\mb{r}(s,t)$ is
governed by the Langevin equation:
\begin{equation}\label{amft2}
\frac{\partial}{\partial t} {r}_\alpha(s,t) = -\int ds^\prime\, \mu^{\alpha}_\text{avg}(s-s^\prime)\frac{\delta U_\text{MFA}}{\delta
r_\alpha(s^\prime,t)} + \xi_\alpha(s,t),
\end{equation}
 where $\xi_\alpha(s,t)$ are stochastic velocities, and hydrodynamic
 effects are included through the pre-averaged anisotropic mobility
 $\mu^{\alpha}_\text{avg}(s-s^\prime)$.  The latter is derived from
 the continuum version of the Rotne-Prager tensor
 $\overleftrightarrow{\bs{\mu}}(s,s^\prime;\mb{x})$~\cite{Harnau1996},
 describing solvent-mediated interactions between two points $s$,
 $s^\prime$ on the contour at spatial separation $\mb{x}$.  If the
 equilibrium probability of finding such a configuration is
 $G(s,s^\prime;\mb{x})$, then the integration $\int
 d^3\mb{x}\,\overleftrightarrow{\bs{\mu}}(s,s^\prime;\mb{x})
 G(s,s^\prime;\mb{x})$ yields a diagonal $3\times 3$ tensor whose
 $\alpha = \parallel, \perp$ components we denote as
 $\mu^{\alpha}_\text{avg}(s-s^\prime)$.  In the absence of
 hydrodynamic effects (a case we will consider as a comparison), the
 free-draining mobility is $\mu^{\alpha}_\text{fd}(s-s^\prime) =
 2a\mu_0 \delta(s-s^\prime)$, where $a$ is a microscopic length scale
 (i.e. the monomer radius), and $\mu_0$ is the Stokes mobility of a
 sphere of radius $a$.  We assume the stochastic velocities
 $\bs{\xi}(s,t)$ are Gaussian, with correlations given by the
 fluctuation-dissipation theorem:
\begin{equation}\label{amft3}
\langle \xi_\alpha(s,t)
 \xi_{\alpha}(s^\prime,t^\prime)\rangle = 2 k_B T
 \delta(t-t^\prime)\mu^\alpha_\text{avg}(s-s^\prime).
\end{equation}
  The Langevin equation, together with boundary conditions at the
  end-points due to the applied force, can be solved through normal
  mode decomposition, yielding all the dynamical quantities which we
  will analyze below.

A reader uninterested in the technical details of the anisotropic MFT
can skip Sec.~\ref{amft:quant}--\ref{amft:dyn} and proceed to the
description of the simulations in Sec.~\ref{bd} and the results in
Sec.~\ref{results}.

\subsection{Quantum solution of the WLC}\label{amft:quant}

The mapping between the WLC and a quantum particle moving on the
surface of a unit sphere can be exploited to calculate exactly many of
the equilibrium properties of the system~\cite{Saito1967}.  Here we follow
a technique similar to Ref.~\citenum{Samuel2002} to numerically evaluate
thermodynamic averages of the WLC to arbitary accuracy.  To compute
all the quantities of interest, we start with a WLC Hamiltonian
augmented by two extra terms:
\begin{equation}\label{qu1}
U_\text{WLC} = \frac{l_p k_B T}{2} \int_0^L ds\, (\partial_s \mb{u}(s))^2 - F \int_0^L ds\,u_z(s) - F_x \int_0^L ds\,u_x(s) - K \int_0^L ds\, u_z^2(s)\,.
\end{equation}
The extra parameters $F_x$ and $K$ will later be set to zero after
taking the appropriate derivatives of the partition function.
Rescaling the contour variable as $\tau = s/l_p$ and factoring out
$k_B T$, we can rewrite $U_\text{WLC}$ in a simpler form:
\begin{equation}\label{qu2}
\beta U_\text{WLC} = \int_0^{\tilde T} d\tau\, \left[\frac{1}{2}(\partial_\tau \mb{u}(\tau))^2 - f u_z(\tau) - f_x u_x(\tau) - k u_z^2(\tau)\right],
\end{equation}
where ${\tilde T}=L/l_p$, $f=\beta l_p F$, $f_x = \beta l_p F_x$, and $k= \beta
l_p K$.  Let us define the propagator $G(\mb{u}_0, \mb{u}_{\tilde T}; {\tilde T})$ as
the path integral over all configurations with initial tangent
$\mb{u}(0) = \mb{u}_0$ and final tangent $\mb{u}({\tilde T}) = \mb{u}_{\tilde T}$:
\begin{equation}\label{qu3}
G(\mb{u}_0, \mb{u}_{\tilde T}; {\tilde T}) = \int_{\mb{u}(0) =
\mb{u}_0}^{\mb{u}({\tilde T}) =
\mb{u}_{\tilde T}} {\cal D}\mb{u} \prod_s
\delta(\mb{u}^2(s)-1)\, \exp(-\beta U_\text{WLC})\,.
\end{equation}
Then for boundary conditions with free end-point tangents the
partition function is given by $Z = (4\pi)^{-2} \int_S d\mb{u}_0\,
d\mb{u}_{\tilde T}\, G(\mb{u}_0, \mb{u}_{\tilde T}; {\tilde T})$, where the integrations are over
the unit sphere $S$.  The quantum Hamiltonian corresponding to $\beta U_\text{WLC}$ is
\begin{equation}\label{qu4}
{\cal H} = -(1/2)\nabla^2 - f \cos \theta -f_x \sin \theta \cos \phi - k \cos^2\theta\,,
\end{equation}
describing a particle on the surface of a unit sphere.  In terms of
the associated quantum eigenvalues $E_n$ and eigenstates $\psi_n(\mb{u})$, the
propagator $G$ is given by:
\begin{equation}\label{qu5}
G(\mb{u}_0, \mb{u}_{\tilde T}; {\tilde T}) = \sum_n e^{-E_n {\tilde T}} \psi_n^\ast(\mb{u}_0) \psi_n(\mb{u}_{\tilde T}) = \sum_{n,l,m,l^\prime,m^\prime} e^{-E_n {\tilde T}} a^\ast_{nl^\prime m^\prime}a_{nlm}Y_{l^\prime m^\prime}^\ast(\mb{u}_0) Y_{lm}(\mb{u}_{\tilde T})\,,
\end{equation}
where in the second step we have expanded out the eigenstates in the
basis of spherical harmonics, $\psi_n(\mb{u}) = \sum_{lm} a_{nlm}
Y_{lm}(\mb{u})$, with coefficients $a_{nlm}$.  For a given $n$, these
coefficients are just the components of the $n$th eigenvector for the
Hamiltonian ${\cal H}$ in the $Y_{lm}$ basis.  Thus to proceed, one
needs the detailed form of this matrix: ${\cal
  H}_{l,m;l^\prime,m^\prime} = {\cal H}^0_{l,m;l^\prime,m^\prime} +
{\cal H}^f_{l,m;l^\prime,m^\prime} + {\cal
  H}^{f_x}_{l,m;l^\prime,m^\prime}+ {\cal
  H}^{k}_{l,m;l^\prime,m^\prime}$.  We list below only the nonzero
elements of each contribution that are relevant to the calculation,
with the symmetry of the matrix implicitly assumed:
\begin{equation}\label{qu6}
\begin{split}
&{\cal H}^{0}_{l,0;l,0} = l(l+1), \quad {\cal H}^{f}_{l,0;l+1,0} = -f (l+1)[(2l+1)(2l+3)]^{-1/2},\\
&{\cal H}^{k}_{l,0;l,0} = -\frac{k (2 l (l+1)-1)}{4 l (l+1)-3}, \quad
{\cal H}^{k}_{l,0;l+2,0} = -\frac{k(l+1) (l+2)}{(2 l+3) \sqrt{4 l^2+12 l+5}},\\
&{\cal H}^{f_x}_{l,m;l+1,m+1}=\frac{1}{2} f_x \sqrt{\frac{(l+m+1) (l+m+2)}{4 l (l+2)+3}}, \quad 
{\cal H}^{f_x}_{l,m;l+1,m-1}= -\frac{1}{2} f_x \sqrt{\frac{(l-m+1) (l-m+2)}{4 l (l+2)+3}} \,.
\end{split}
\end{equation}
The eigenvectors $a_{nlm}$ and eigenvalues $E_n$ can be readily
calculated numerically by truncating the infinite matrix ${\cal
  H}_{l,m;l^\prime,m^\prime}$ to a finite size (with cutoff chosen
large enough to get the desired precision).  The partition function
$Z$ can then be written as:
\begin{equation}\label{qu7}
Z = \frac{1}{(4\pi)^2} \int_S d\mb{u}_0\, d\mb{u}_{\tilde T}\, \sum_{n,l,m,l^\prime,m^\prime} e^{-E_n {\tilde T}} a^\ast_{nl^\prime m^\prime}a_{nlm}Y_{l^\prime m^\prime}^\ast(\mb{u}_0) Y_{lm}(\mb{u}_{\tilde T}) = \frac{1}{4\pi} \sum_n e^{-E_n {\tilde T}} a^\ast_{n00}a_{n00}.
\end{equation}
Most of the thermodynamic averages used in the anisotropic MFT can be
directly derived from Z:
\begin{equation}\label{qu8}
\begin{split}
\langle R_\parallel \rangle &= \left. l_p\frac{\partial}{\partial f}\log Z\right|_{f_x=0,k=0}, \qquad \int_0^L ds\,\langle u_\parallel^2(s) \rangle = L-\int_0^L ds\,\langle u_\perp^2(s) \rangle = \left.l_p \frac{\partial}{\partial k}\log Z\right|_{f_x=0,k=0},\\
\langle \delta R_\parallel^2 \rangle &= \left. l_p^2\frac{\partial^2}{\partial f^2}\log Z\right|_{f_x=0,k=0},\qquad \langle \delta R_\perp^2 \rangle = \left. 2l_p^2\frac{\partial^2}{\partial f_x^2}\log Z\right|_{f_x=0,k=0}.
\end{split}
\end{equation}
The end-point averages are calculated similarly:
\begin{equation}\label{qu9}
\begin{split}
\langle u_{0\parallel}^2(0) + u_{0\parallel}^2(L) \rangle = 2- \langle u_{0\perp}^2(0) + u_{0\perp}^2(L) \rangle &= \frac{2}{(4\pi)^2} \int_S d\mb{u}_0\, d\mb{u}_{\tilde T}\,  u^2_{0\parallel} G(\mb{u}_0, \mb{u}_{\tilde T}; {\tilde T})\\
& = \sum_n e^{-E_n {\tilde T}} a^\ast_{n00} \left(\frac{a_{n00}}{6\pi} + \frac{a_{n20}}{3\sqrt{5}\pi} \right).
\end{split}
\end{equation}

\subsection{Calculating the parameters of the anisotropic MFT}\label{amft:par}

\begin{figure}[!t]
\includegraphics[scale=1]{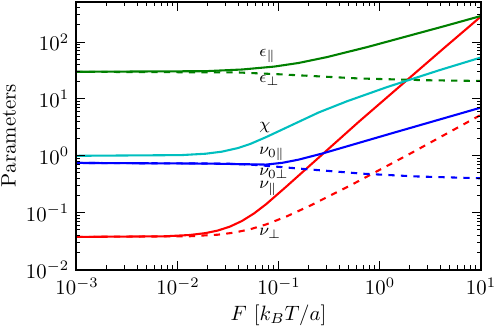}
\caption{Parameters of the anisotropic MFT Hamiltonian $U$ as a
  function of force $F$ for a chain with $L=100a$, $l_p=20a$, derived
  as described in Sec.~\ref{amft:quant} and \ref{amft:par} in order to
  reproduce exact equilibrium averages of the WLC. To plot all
  parameters in one graph, the units for the $y$-axis are as follows:
  $k_BT a$ for $\epsilon_\alpha$, $k_B T/a$ for $\nu_\alpha$, $k_BT$
  for $\nu_{0\alpha}$, $\alpha = \parallel,\perp$.  The force
  rescaling parameter $\chi$ is dimensionless.  The scaling forms of
  all these parameters in the small and large force limits are
 shown at the end of Sec.~\ref{amft:par}.}\label{fig1}
\end{figure}

With free end-point tangent boundary conditions, the partition
function
\begin{equation}\label{cal1}
Z_\text{MFA} = (4\pi)^{-2} \int_S d\mb{u}_0\, d\mb{u}_L\, \int_{\mb{u}(0) =
\mb{u}_0}^{\mb{u}(L) =
\mb{u}_L} {\cal D}\mb{u} \exp(-\beta U_\text{MFA})
\end{equation}
 corresponding to the anisotropic MFT Hamiltonian
$U_\text{MFA}$, Eq.~\eqref{amft1}, can be evaluated analytically:
\begin{equation}\label{cal2}
\begin{split}
Z_\text{MFA} &= \frac{2 \sqrt{2} \pi ^{3/2} \beta^{-1} \epsilon_\perp \omega_\perp}{\left(\epsilon_\perp^2 \omega_\perp^2+4 \nu_{0\perp}^2\right) \sinh (L
   \omega_\perp)+4 \epsilon_\perp \nu_{0\perp} \omega_\perp \cosh (L \omega_\perp)}\\
&\qquad\qquad \cdot
\sqrt{\frac{\beta^{-1} \epsilon_\parallel \omega_\parallel }{\left(\epsilon_\parallel^2 \omega_\parallel ^2+4 \nu_{0\parallel}^2\right) \sinh (L \omega_\parallel )+4 \epsilon_\parallel \nu_{0\parallel} \omega_\parallel  \cosh (L
   \omega_\parallel )}}\\
&\qquad\qquad \cdot \exp \left(\frac{\beta \chi^2 F^2 \left(\sinh \left(\frac{L \omega_\parallel }{2}\right) \left(\epsilon_\parallel L \omega_\parallel ^2-4 \nu_{0\parallel}\right)+2 L \nu_{0\parallel} \omega_\parallel  \cosh
   \left(\frac{L \omega_\parallel }{2}\right)\right)}{2 \epsilon_\parallel \omega_\parallel ^3 \left(\epsilon_\parallel \omega_\parallel  \sinh \left(\frac{L \omega_\parallel }{2}\right)+2 \nu_{0\parallel}
   \cosh \left(\frac{L \omega_\parallel }{2}\right)\right)}\right)\,,
\end{split}
\end{equation}
where $\omega_\alpha \equiv \sqrt{2\nu_\alpha/\epsilon_\alpha}$,
$\alpha=\parallel,\perp$.  Similarly one can extract analytic
expressions for all seven of the thermodynamic averages used in the
fitting of the MFT parameters:
\begin{equation}\label{cal3}
\begin{split}
&\langle R_\parallel \rangle = (\beta\chi)^{-1} \frac{\partial}{\partial F}\log Z_\text{MFA}, \qquad \int_0^L ds\,\langle u_\parallel^2(s) \rangle = - \beta^{-1}\frac{\partial}{\partial \nu_\parallel}\log Z_\text{MFA},\\
& \int_0^L ds\,\langle u_\perp^2(s) \rangle = - \beta^{-1} \frac{\partial}{\partial \nu_\perp}\log Z_\text{MFA}, \qquad \langle \delta R_\parallel^2 \rangle = (\beta\chi)^{-2} \frac{\partial^2}{\partial F^2}\log Z_\text{MFA},\\
&\langle \delta R_\perp^2 \rangle = \frac{2}{\beta \epsilon_\perp \omega_\perp^3} \left(L \omega_\perp-\frac{4 \nu_{0\perp}}{2 \nu_{0\perp} \coth \left(\frac{L \omega_\perp}{2}\right)+\epsilon_\perp
   \omega_\perp}\right),\\
&\langle u_{0\parallel}^2(0) + u_{0\parallel}^2(L) \rangle = - \beta^{-1}\frac{\partial}{\partial \nu_{0\parallel}}   \log Z_\text{MFA},\qquad \langle u_{0\perp}^2(0) + u_{0\perp}^2(L) \rangle = - \beta^{-1}\frac{\partial}{\partial \nu_{0\perp}}\log Z_\text{MFA}. 
\end{split}
\end{equation}
By setting these expressions equal to the exact WLC results calculated
from the quantum approach, Eqs.~\eqref{qu8} and \eqref{qu9}, one
obtains a system of seven coupled equations that can be solved
numerically for a given $L$, $l_p$, and $F$, yielding the seven
Hamiltonian parameters: $\epsilon_\parallel$, $\epsilon_\perp$,
$\nu_\parallel$, $\nu_\perp$, $\nu_{0\parallel}$, $\nu_{0\perp}$, and
$\chi$.  \ref{fig1} shows a set of solutions for $L=100a$, $l_p=20a$,
and varying $F$.  In the small force regime, $F \ll k_B T/l_p$, where
stretching is negligible, the parameters converge to the same values
as in the isotropic MFT: $\nu_\parallel = \nu_\perp = 3k_BT/4l_p$,
$\nu_{0\parallel} = \nu_{0\perp} = 3k_BT/4$, $\epsilon_\parallel =
\epsilon_\perp = 3k_BTl_p/2$ \cite{Winkler1994,Ha1995}.  In the
opposite limit of large force, $F \gg k_B T/l_p$, we clearly see
symmetry breaking between the tangential and perpendicular parameters,
and the model becomes distinctly anisotropic.  In this regime the
parameters scale like: $\nu_\parallel = 2 (l_p F^3/k_B T)^{1/2}$,
$\nu_\perp = F/2$, $\nu_{0\parallel} = (l_p F k_B T)^{1/2}/2$,
$\nu_{0\perp} \approx 0.38 k_BT$, $\epsilon_\parallel = (l_p^3 F k_B
T)^{1/2}$, $\epsilon_\perp = l_p k_BT$, and $\chi = 4(l_p
F/k_BT)^{1/2}$.  As discussed in Sec.~\ref{results:wbacomp}, when the
chain approaches full extension with $F \to \infty$, the transverse
part of the MFT Hamiltonian converges to the correct WBA asymptotic
limit.  The longitudinal part does not have the WBA limiting behavior,
but this breakdown is expected, since a Gaussian model cannot describe
the longitudinal dynamics of a nearly rigid rod.

\subsection{Anisotropic MFT dynamical theory}\label{amft:dyn}

Here we adapt the dynamical theory used successfully for the isotropic
MFT~\cite{Harnau1996, Hinczewski2009} to the anisotropic case of a
chain under tension.  The general hydrodynamic pre-averaging approach
is similar to that used for the Zimm model~\cite{Zimm1956,DoiEdwards}.
The time evolution of the chain $\mb{r}(s,t)$ follows the Langevin
equation:
\begin{equation}\label{d1}
\begin{split}
&\frac{\partial}{\partial t}\mb{r}(s,t) = -\int_{0}^{L} ds^\prime\, \overleftrightarrow{\bs{\mu}}\left(s,s^\prime;\mb{r}(s,t)-\mb{r}(s^\prime,t)\right) \frac{\delta U_\text{MFA}}{\delta \mb{r}(s^\prime,t)}  + \bs{\xi}(s,t)\,.
\end{split}
\end{equation}
Here the $\bs{\xi}(s,t)$ is the stochastic contribution, and
$\overleftrightarrow{\bs{\mu}}(s,s^\prime;\mb{x})$ is the continuum
version of the Rotne-Prager tensor~\cite{Harnau1996},
\begin{equation}\label{d2}
\begin{split}
\overleftrightarrow{\bs{\mu}}(s,s^\prime;\mb{x}) =& \quad 2a\mu_0 \delta(s-s^\prime)\overleftrightarrow{\mb{1}} + \Theta(x-2a)\left(\frac{1}{8\pi\eta x} \left[\overleftrightarrow{\mb{1}} +\frac{\mb{x}\otimes\mb{x}}{x^2}\right] + \frac{a^2}{4\pi\eta x^3}\left[ \frac{\overleftrightarrow{\mb{1}}}{3} - \frac{\mb{x}\otimes\mb{x}}{x^2}\right]\right)\,,
\end{split}
\end{equation}
describing long-range hydrodynamic interactions between two points at
$s$ and $s^\prime$ on the chain contour, separated by a spatial
distance $\mb{x}$.  The microscopic length scale $a$ corresponds to
the monomer radius, $\eta$ is the viscosity of water, $\mu_0 = 1/6\pi
\eta a$ is the Stokes mobility of a sphere of radius $a$, and the
$\Theta$ function excludes unphysical configurations (overlap between
monomers).

\begin{figure}[!t]
\includegraphics[scale=1]{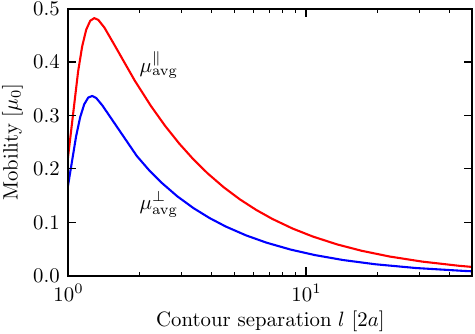}
\caption{Pre-averaged mobilities $\mu^\parallel_\text{avg}(l)$ and
  $\mu^\perp_\text{avg}(l)$ as a function of contour separation $l$
  for a chain with $L=100a$, $l_p = 20a$, and $F=1.0$ $k_B
  T/a$.}\label{fig2}
\end{figure}

Eq.~\eqref{d1} cannot be solved directly, because the hydrodynamic
tensor depends on the exact configuration of the chain at time $t$, so
we employ the pre-averaging approximation: replacing
$\overleftrightarrow{\bs{\mu}}(s,s^\prime;\mb{r}(s,t)-\mb{r}(s^\prime,t))$
with an average over all equilibrium configurations,
$\overleftrightarrow{\bs{\mu}}_\text{avg}(s,s^\prime)$:
\begin{equation}\label{d3}
\begin{split}
&\overleftrightarrow{\bs{\mu}}_\text{avg}(s,s^\prime) = \int d^3\mb{x}\, \overleftrightarrow{\bs{\mu}}(s,s^\prime;\mb{x}) G(s,s^\prime;\mb{x})\,,
\end{split}
\end{equation}
where $ G(s,s^\prime;\mb{x})$ is the equilibrium probability of
finding two points at $s$ and $s^\prime$ with spatial separation
$\mb{x}$.  For the anisotropic Hamiltonian $U_\text{MFA}$ this
probability takes the form:
\begin{equation}\label{d4}
G(s,s^\prime;\mb{x}) = \frac{3}{2\pi \sigma_\perp(s-s^\prime)} \left( \frac{3}{2\pi \sigma_\parallel(s-s^\prime)}\right)^{1/2} \exp\left(-\frac{3 x_\perp^2}{ 2 \sigma_\perp(s-s^\prime)} -\frac{3 (x_\parallel - \chi F |s-s^\prime|/2\nu_\parallel)^2}{2 \sigma_\parallel(s-s^\prime)} \right)\,, 
\end{equation}
where $\sigma_\alpha(l) \equiv (3 (|l| \omega_\alpha +\exp(-|l|
\omega_\alpha)-1)/\beta \epsilon_\alpha \omega_\alpha^3$.  In deriving $G$
we have assumed a large chain length $L$, which simplifies the
resulting analytical expression.  Plugging Eq.~\eqref{d4} into
Eq.~\eqref{d3} leads to:
\begin{equation}\label{d5}
\overleftrightarrow{\bs{\mu}}_\text{avg}(s,s^\prime) = \begin{pmatrix} \mu^\perp_\text{avg}(s-s^\prime) & 0 & 0 \\
0 & \mu^\perp_\text{avg}(s-s^\prime) & 0 \\
0 & 0 & \mu^\parallel_\text{avg}(s-s^\prime) \end{pmatrix},
\end{equation}
where the anisotropic mobilities $\mu^\alpha_\text{avg}$ can be
written in terms of integrals over coordinates $x = \sqrt{x^2_\perp +
  x^2_\parallel}$ and $\zeta = x_\parallel / x$:
\begin{equation}\label{d6}
\begin{split}
\mu_\text{avg}^\parallel(l) =& 2a\mu_0 \delta(l) + \frac{3^{3/2}\Theta(l - 2a)\mu_0}{(2\pi)^{3/2} \sigma_\perp(l) \sqrt{\sigma_\parallel(l)}} \int_{2a}^\infty dx\, \int_{-1}^1 d\zeta\,\left(\frac{\pi -3 \pi  \zeta^2}{x}+\frac{3}{2} \pi  \left(\zeta^2+1\right) x\right)\\
& \hspace{2.5in}\cdot \exp \left(\frac{3
   \left(\zeta^2-1\right) x^2}{2 \sigma_\perp(l)}-\frac{3 ( \zeta x - \chi F l/2\nu_\parallel)^2}{2 \sigma_\parallel(l) }\right),\\
\mu_\text{avg}^\perp(l) &= 2a\mu_0 \delta(l) + \frac{3^{3/2}\Theta(l - 2a)\mu_0}{(2\pi)^{3/2} \sigma_\perp(l) \sqrt{\sigma_\parallel(l)}} \int_{2a}^\infty dx\, \int_{-1}^1 d\zeta\, \frac{\pi  \left(-3 \zeta^2 \left(x^2-2\right)+9 x^2-2\right)}{4 x} \\
& \hspace{2.5in} \cdot \exp \left(\frac{3
   \left(\zeta^2-1\right) x^2}{2 \sigma_\perp(l)}-\frac{3 ( \zeta x - \chi F l/2\nu_\parallel)^2}{2 \sigma_\parallel(l) }\right).
\end{split}
\end{equation}
These integrals are evaluated numerically to obtain the mobilities as
a function of contour distance $l$.  In \ref{fig2} we show the
results for $L=100a$, $l_p=20a$, and $F=1.0$ $k_BT/a$.  Note that the
mobility parallel to the stretching direction is enhanced relative to
the transverse component, as we expect for an extended chain.

The pre-averaged version of the Langevin equation can now be written as:
\begin{equation}\label{d7}
\begin{split}
\frac{\partial}{\partial t}r_\alpha(s,t) = -\int_{0}^{L} ds^\prime\,\mu^\alpha_\text{avg}(s-s^\prime)\frac{\delta U_\text{MFA}}{\delta r_\alpha(s^\prime,t)} + \xi_\alpha(s,t),
\end{split}
\end{equation}
for $\alpha = \parallel, \perp$.  The $\bs{\xi}(s,t)$ are Gaussian
random vectors, whose components have correlations given by the
fluctuation-dissipation theorem:
\begin{equation}\label{d8}
\langle \xi_\alpha(s,t)\xi_{\alpha}(s^\prime,t^\prime)\rangle = 2 k_B T\delta(t-t^\prime)\mu^\alpha_\text{avg}(s-s^\prime).
\end{equation}
Plugging in the form of $U_\text{MFA}$, the internal force term in the
Langevin equation becomes
\begin{equation}\label{d9}
\frac{\delta U_\text{MFA}}{\delta r_\alpha(s^\prime,t)} =\epsilon_\alpha \frac{\partial^4}{\partial {s^\prime}^4}r_\alpha(s^\prime,t)-2\nu_\alpha\frac{\partial^2}{\partial {s^\prime}^2} r_\alpha(s^\prime,t) \equiv \hat{O}^\alpha_{s^\prime} r_\alpha(s^\prime,t), 
\end{equation}
where we have introduced the differential operator
$\hat{O}^\alpha_{s^\prime}$.  To complete the dynamical theory, we
must specify the boundary conditions at the chain ends:
\begin{equation}\label{d10}
\begin{split}
-\epsilon_\alpha \frac{\partial^3}{\partial s^3}r_\alpha(0,t) +2\nu_\alpha \frac{\partial}{\partial s} r_\alpha(0,t) &= \chi F\delta_{\alpha,\parallel},\qquad
-\epsilon_\alpha \frac{\partial^3}{\partial s^3}r_\alpha(L,t) +2\nu_\alpha \frac{\partial}{\partial s} r_\alpha(L,t) = \chi F\delta_{\alpha,\parallel},\\
\epsilon_\alpha \frac{\partial^2}{\partial s^2}r_\alpha(0,t) - 2\nu_{0\alpha} \frac{\partial}{\partial s} r_\alpha(0,t) &= 0,\qquad
-\epsilon_\alpha \frac{\partial^2}{\partial s^2}r_\alpha(L,t) - 2\nu_{0\alpha} \frac{\partial}{\partial s} r_\alpha(L,t) = 0.
\end{split}
\end{equation}
The first two represent the force applied at the ends, while the
second two the absence of torque.  To properly deal with the boundary
conditions for $F\ne 0$, we write $\mb{r}(s,t)$ in the following way:
$r_\alpha(s,t) = \tilde{r}_\alpha(s,t) + \chi F \delta_{\alpha,\parallel}
\phi(s)$, where $\tilde{\mb{r}}(s,t)$ satisfies the homogeneous ($F=0$)
version of Eq.~\eqref{d10}, while $\phi(s)$ is chosen such that the
total function $\mb{r}(s,t)$ satisfies the full boundary requirements.
The resulting form for $\phi(s)$ is:
\begin{equation}\label{d11}
\phi(s) = \frac{Ls}{2(L\nu_\parallel+2\nu_{0\parallel})} + \frac{L \nu_{0\parallel} s^2}{2\epsilon_\parallel(L\nu_\parallel+2\nu_{0\parallel})} - \frac{\nu_{0\parallel}s^3}{3\epsilon_\parallel(L\nu_\parallel+2\nu_{0\parallel})}.
\end{equation}

We now proceed to transform the Langevin equation into matrix form,
which will allow us to solve it through numerical diagonalization. Let
us assume $\xi_\alpha(s,t)$ satisfies similar boundary conditions to
$\tilde{r}_\alpha(s,t)$, and expand both functions in normal modes
$\psi^\alpha_n(s)$, with amplitudes $p_{\alpha n}(t)$ and
$q_{\alpha n}(t)$ respectively:
\begin{equation}\label{d12}
\tilde{r}_\alpha(s,t) = \sum_{n=0}^\infty p_{\alpha n}(t) \psi^\alpha_n(s), \quad \xi_\alpha(s,t) = \sum_{n=0}^\infty q_{\alpha n}(t) \psi^\alpha_n(s)\,.
\end{equation}
The normal modes $\psi^\alpha_{n}(s)$ are chosen to be eigenfunctions
of the differential operator $\hat{O}^\alpha_s$, satisfying
$\hat{O}^\alpha_s \psi^\alpha_{n}(s) = \lambda_{\alpha n} \psi^\alpha_{n}(s)$
for eigenvalues $\lambda_{\alpha n}$.  These eigenfunctions take the form~\cite{Harnau1996}:
\begin{equation}\label{d13}
\begin{split}
\psi^\alpha_0(s) &= \sqrt{\frac{1}{L}}\,,\\
\psi^\alpha_n(s) &= \sqrt{\frac{C_{\alpha n}}{L}}\left( K_{\alpha n} \frac{\sin K_{\alpha n} (s-L/2)}{\cos K_{\alpha n} L/2} + G_{\alpha n} \frac{\sinh G_{\alpha n} (s-L/2)}{\cosh G_{\alpha n} L/2}\right)\,, \quad n\: \text{odd},\\
\psi^\alpha_n(s) &= \sqrt{\frac{C_{\alpha n}}{L}}\left( -K_{\alpha n} \frac{\cos K_{\alpha n} (s-L/2)}{\sin K_{\alpha n} L/2} + G_{\alpha n} \frac{\cosh G_{\alpha n} (s-L/2)}{\sinh G_{\alpha n} L/2}\right)\,, \quad n\: \text{even},
\end{split}
\end{equation}
where  
\begin{equation}\label{d14}
G_{\alpha n}^2 - K_{\alpha n}^2 = 2\nu_\alpha/\epsilon_\alpha\,, \quad \lambda^\alpha_0=0\,,\quad \lambda_{\alpha n} = \epsilon_\alpha K_{\alpha n}^4 + 2\nu_\alpha G_{\alpha n}^2\,.
\end{equation} 
The eigenfunctions obey the boundary conditions in the $F=0$ version
of Eq.~\eqref{d10}, which fixes $\lambda_{\alpha n}$, and hence the
constants $K_{\alpha n}$ and $G_{\alpha n}$, while the $C_{\alpha n}$
are normalization coefficients.  The boundary conditions lead to a
single transcendental equation for $\lambda_{\alpha n}$, whose
solutions can be found easily using a standard numerical root finding
algorithm.  Plugging the normal mode expansions into the Langevin
equation, and exploiting the orthonormality of the $\psi^\alpha_n$,
Eqs.~\eqref{d7}-\eqref{d8} become:
\begin{equation}\label{d15}
\begin{split}
\frac{\partial}{\partial t}p_{\alpha n}(t) &= - \sum_{m=0}^\infty H^\alpha_{nm} \lambda_{\alpha m} p_{\alpha m}(t) + \chi F w_n \delta_{\alpha,\parallel} + q_{\alpha n}(t)\,,\\
 \langle q_{\alpha n}(t) q_{\alpha m}(t^\prime) \rangle &= 2 k_B T \delta(t-t^\prime)H^\alpha_{nm},
\end{split}
\end{equation}
where
\begin{equation}\label{d16}
\begin{split}
H^\alpha_{nm} = \int_{0}^{L}ds \int_{0}^{L}ds^\prime\,\psi^\alpha_n(s)\mu^\alpha_\text{avg}(s-s^\prime)\psi^\alpha_m(s^\prime)\,,\\
w_n = \int_{0}^{L}ds \int_{0}^{L}ds^\prime\, \psi^\parallel_n(s) \mu^\parallel_\text{avg}(s-s^\prime) \frac{2\nu_\parallel \nu_{0\parallel} (L - 2s^\prime)}{\epsilon_\parallel (L \nu_\parallel + 2 \nu_{0\parallel})}.
\end{split}
\end{equation}
Both $H^\alpha_{nm}$ and $w_n$ can be evaluated through numerical
integration.  In order to make solving these equations feasible, we
introduce a high-frequency cutoff $M$ on the mode number, keeping only
the slowest-relaxing modes $n = 0,\ldots,M-1$ (the $M$ modes with
smallest $\lambda_{\alpha n}$), whose hydrodynamic interactions are
described by the leading $M \times M$ sub-blocks of the matrices
$H^\alpha$.  Following Ref.~\citenum{Hinczewski2009} we set $M =
L/8a$, which provides good agreement at short times with Brownian
dynamics simulations of bead-spring chains with monomer radius $a$.
For longer times, where the polymer motion is on length scales much
larger than $a$, the dynamical results are insensitive to the precise
value of the cutoff.  Overall the numerical cost of evaluating all the
quantities in the theory is quite small, and can be accomplished on
the order of minutes for $M=12$, the cutoff size used for the $L=100a$
chains studied here.

The final step in simplifying the dynamical theory is diagonalization.
Let $J^\alpha$ be the $M \times M$ matrix with elements $J^\alpha_{nm}
= H^\alpha_{nm} \lambda_{\alpha m}$, $\Lambda_{\alpha n}$ be the
eigenvalues of $J^\alpha$, and $C^\alpha$ the matrix diagonalizing
$J^\alpha$: $[C^\alpha J^\alpha (C^\alpha)^{-1}]_{nm} =
\Lambda_{\alpha n} \delta_{nm}$.  The $\Lambda_{\alpha n}$ are assumed
ordered from smallest to largest with increasing $n$.  Assuming
non-degenerate eigenvalues $\Lambda_{\alpha n}$, the matrix $C^\alpha$
also diagonalizes $H^\alpha$ through the congruent transformation
$[C^\alpha H^\alpha (C^\alpha)^T]_{nm} = \Theta_{\alpha n}
\delta_{nm}$, defining parameters $\Theta_{\alpha n}$ \cite{Zimm1956}.
The diagonal version of Eq.~\eqref{d15} then reads:
\begin{equation}\label{d17}
\begin{split}
\frac{\partial}{\partial t}P_{\alpha n}(t) &= - \Lambda_{\alpha n} P_{\alpha n}(t) + \chi F W_n \delta_{\alpha,\parallel} + Q_{\alpha n}(t)\,,\\
 \langle Q_{\alpha n}(t) Q_{\alpha m}(t^\prime) \rangle &= 2 k_B T \delta(t-t^\prime) \delta_{m,n} \Theta_{\alpha n},
\end{split}
\end{equation}
where
\begin{equation}\label{d18}
\begin{split}
P_{\alpha n}(t) &= \sum_{m=0}^{M-1} C^\alpha_{nm} p_{\alpha m}(t),\quad Q_{\alpha n}(t) = \sum_{m=0}^{M-1} C^\alpha_{nm} q_{\alpha m}(t), \quad W_n = \sum_{m=0}^{M-1} C^\parallel_{nm} w_m, 
\end{split}
\end{equation} 
and $r_\alpha(s,t) = \sum_n P_{\alpha n}(t) \Psi^\alpha_n(s)
+ \chi F\delta_{\alpha,\parallel} \phi(s)$ with modified normal modes
$\Psi^\alpha_n (s) = \sum_m \psi^\alpha_m(s)
[(C^{\alpha})^{-1}]_{mn}$.  Using Eq.~\eqref{d17} it now becomes
possible to solve for a variety of dynamical observables.  For
example, the result for the MSD of a chain
end-point is:
\begin{equation}\label{eq:d19}
\begin{split}
\Delta^\text{end}_\alpha(t) \equiv \langle \left(r_\alpha(L,t)-r_\alpha(L,0)\right)^2 \rangle &= 2 k_B T \left[\Theta_{\alpha 0} (\Psi^\alpha_0(L))^2 t + \sum_{n>0} \frac{\Theta_{\alpha n}}{\Lambda_{\alpha n}}(1-\exp(-\Lambda_{\alpha n} t)) (\Psi^\alpha_n(L))^2 \right]\\
&= 2 D_\alpha t + 2k_B T \sum_{n>0} A^\text{end}_{\alpha n}(1-\exp(-\Lambda_{\alpha n} t)),
\end{split}
\end{equation}
where we have introduced the center-of-mass diffusion constant
$D_\alpha = k_B T \Theta_{\alpha 0} (\Psi^\alpha_0(L))^2$, and
coefficients $A^\text{end}_{\alpha n} = \Theta_{\alpha
    n} (\Psi^\alpha_n(L))^2/\Lambda_{\alpha n}$.  Similarly, for the MSD of the end-to-end vector,
\begin{equation}\label{eq:d20}
\begin{split}
\Delta^\text{ee}_\alpha(t) \equiv \langle \left(R_\alpha(t)-R_\alpha(0)\right)^2 \rangle &= 2k_B T \sum_{n>0} \frac{\Theta_{\alpha n}}{\Lambda_{\alpha n}}(1-\exp(-\Lambda_{\alpha n} t)) (\Psi^\alpha_n(L)-\Psi^\alpha_n(0))^2\\
&= 2k_B T \sum_{n>0} A^\text{ee}_{\alpha n}(1-\exp(-\Lambda_{\alpha n} t)),
\end{split}
\end{equation}
where $A^\text{ee}_{\alpha n} = \Theta_{\alpha n}
(\Psi^\alpha_n(L)-\Psi^\alpha_n(0))^2/\Lambda_{\alpha n}$.  One can
see from the form of Eqs.~\eqref{eq:d19}-\eqref{eq:d20} that the
eigenvalues $\Lambda_{\alpha n}$ correspond to inverse relaxation
times $\tau_{\alpha n}^{-1} \equiv \Lambda_{\alpha n}$.

\section{Brownian dynamics simulations}\label{bd}

For the BD simulations~\cite{Ermak1978} used to test the mean-field
theory, the chain consists of $N$ beads of radius $a$ (contour length
$L= 2aN$) whose positions $\mb{r}_i(t)$ are governed by the discrete
Langevin equation:
\begin{equation}\label{bd1}
\frac{d\mb{r}_i(t)}{dt} = \sum_{j=1}^N \overleftrightarrow{\bs{\mu}}_{ij} \cdot \left(-\frac{\partial U_\text{BD}(\mb{r}_1,\ldots,\mb{r}_N)}{\partial \mb{r}_j} \right)+\bs{\xi}_i(t)\,.
\end{equation}
Long-range hydrodynamic interactions between monomers are included
through the Rotne-Prager \cite{Rotne1969} mobility matrix
$\overleftrightarrow{\bs{\mu}}_{ij}$, which is a discrete version of
Eq.~\eqref{d2}:
\begin{equation}\label{bd2}
\begin{split}
\overleftrightarrow{\bs{\mu}}_{ij} =& \mu_0 \delta_{i,j} \overleftrightarrow{\mb{1}} + (1-\delta_{i,j})\left(\frac{1}{8\pi\eta r_{ij}} \left[\overleftrightarrow{\mb{1}} +\frac{\mb{r}_{ij}\otimes\mb{r}_{ij}}{r_{ij}^2}\right] + \frac{a^2}{4\pi\eta r_{ij}^3}\left[ \frac{\overleftrightarrow{\mb{1}}}{3} - \frac{\mb{r}_{ij}\otimes\mb{r}_{ij}}{r_{ij}^2}\right]\right)\,,
\end{split}
\end{equation}
where $\mb{r}_{ij} \equiv \mb{r}_i - \mb{r}_j$.  This matrix also
determines correlations for the Gaussian stochastic velocities
$\bs{\xi}_i(t)$ according to the fluctuation-dissipation theorem:
\begin{equation}\label{bd3}
\langle \bs{\xi}_i(t) \otimes \bs{\xi}_j(t^\prime) \rangle = 2 k_B T
\overleftrightarrow{\bs{\mu}}_{ij} \delta(t-t^\prime)\,.
\end{equation}

The elastic potential of the chain $U_\text{BD} = U_\text{ben} +
U_\text{str} + U_\text{LJ} + U_\text{ext}$ consists of four parts: (i)
a bending energy $U_\text{ben} = (\epsilon_\text{BD}/2a)\sum_{i}
(1-\cos\theta_i)$, where $\theta_i$ is the angle between two adjacent
bonds, and $\epsilon_\text{BD}$ is related to the persistence length
$l_p$ as $\epsilon_\text{BD} = l_p k_B T$; (ii) a harmonic stretching
term $U_\text{str} = (\gamma/4a) \sum_i \left(r_{i+1,i}-2a\right)^2$
where inextensibility is enforced through a large modulus $\gamma =
2000 k_B T /a$; for a recent discussion of the effects of varying
stretching modulus strength and the competition between bending and
stretching fluctuations, see Ref.~\cite{vonHansen2011} ; (iii) a
truncated Lennard-Jones interaction $U_\text{LJ} = \omega \sum_{i < j}
\Theta(2a-r_{ij})[(2a/r_{ij})^{12}-2 (2a/r_{ij})^{6} +1]$ with $\omega
= 3k_B T$; (iv) an external force $F$ along the z direction,
$U_\text{ext} = -F \hat{\mb{z}} \cdot (\mb{r}_N - \mb{r}_1)$.

In the numerical implementation of Eq.~\eqref{bd1}, the Langevin time
step is $\tau = 3 \times 10^{-4}$ $a^2/(k_BT\mu_0)$, where $\mu_0$ is
the Stokes mobility of a monomer, and a typical simulation lasts $\sim
10^8 - 10^9$ steps.  Data is collected every $10^2 - 10^3$ steps, and
averages for the dynamical quantities discussed below are based on
5-25 independent runs.

\section{Results and Discussion}\label{results}

\subsection{Comparison with BD simulations}\label{results:bdcomp}

\begin{figure}[!t]
\includegraphics*[scale=1]{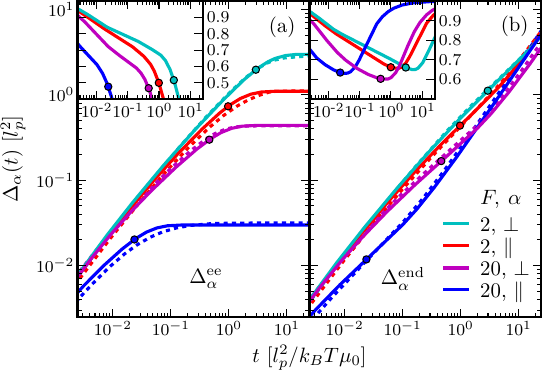}
\caption{MSD functions for a polymer with $L=100a$, $L/l_p = 5$: (a)
  the chain end-point MSD $\Delta^\text{end}_\alpha(t) = \langle
  (r_\alpha(L,t) -r_\alpha(L,0))^2\rangle$; (b) the end-to-end vector
  MSD $\Delta^\text{ee}_\alpha(t) = \langle (R_\alpha(t) -
  R_\alpha(0))^2 \rangle$, where $R_\alpha(t) = r_\alpha(L,t) -
  r_\alpha(0,t)$.  Solid lines are the anisotropic MFT results, while
  dashed lines are taken from BD simulations.  In all cases long-range
  hydrodynamic interactions are taken into account, and results are
  given for $\alpha=\perp,\parallel$ at two forces, $F=2$ and $20\:
  k_B T/l_p$.  Filled circles mark the relaxation times $\tau_{\alpha
    1}$, derived from the MFT, while the insets show the local slopes
  of MFT curves in the log-log plots.}
\label{msdalp}
\end{figure}

To validate the anisotropic MFT, we compared the theoretical results
to BD simulations of a bead-spring worm-like chain.  We focused on two
types of dynamical quantities, both of which are in principle
experimentally accessible (i.e. in an optical tweezer setup): (i) MSD
functions related to the polymer end-points; (ii) the associated
linear response functions, connected to the MSD through the fluctuation-dissipation theorem.

\ref{msdalp} shows MSD results for a representative semiflexible
polymer, with $L=100a$ and $L/l_p = 5$.  For each direction $\alpha$,
the MSD of the end-to-end vector, $\Delta^\text{ee}_\alpha(t) \equiv
\langle (R_\alpha(t) - R_\alpha(0))^2 \rangle$ [\ref{msdalp}(a)], and
an end-point of the chain, $\Delta^\text{end}_\alpha(t) \equiv \langle
(r_\alpha(L,t) - r_\alpha(L,0))^2 \rangle$ [\ref{msdalp}(b)], is
depicted at two different forces $F$.  There is excellent quantitative
agreement with the BD simulations (dashed curves), with the maximum
errors $\approx 10\%$ for the $\perp$ and $\approx 20\%$ for the
$\parallel$ results in the time ranges shown.  The biggest
discrepancies occur at short times for the $\parallel$ component with
$F=20 \: k_BT/l_p$, where the length scale of the motion is comparable
to the bead size, and we expect the discrete BD chain to deviate from
continuum MFT behavior.

The close agreement is all the more remarkable since the MSD shows a
complex crossover behavior.  Asymptotic WBA scaling theory for the
transverse dynamics predicts that for $t \ll \tau_{\perp 1}$, the
longest relaxation time in the $\perp$ direction, there are two
regimes separated by the crossover time $t^\ast = 2l_p k_B T / 3F^2
\mu_0 a$~\cite{Granek1997}: a stiffness-dominated regime at $t \ll
t^\ast$, with MSD $\propto t^{3/4}$, and a force-dominated regime at
$\tau_{\perp 1} \gg t \gg t^\ast$, with a slower scaling $\propto
t^{1/2}$.  The insets of \ref{msdalp}(a)-(b) show the local slopes of
the log-log MSD plots, $d\log \Delta_\alpha/d\log t$, with times
$\tau_{\perp 1}$ calculated from the MFT marked by dots.  With
increasing $F$, we do indeed find the local slope is reduced, but the
dynamic scaling is modified by two important effects: (i) the slow
crossover to center-of-mass motion at times $t \gg \tau_{\perp 1}$,
where the slopes of $\Delta_\alpha^\text{end}$ and
$\Delta_\alpha^\text{ee}$ approach 1 and 0 respectively; (ii)
logarithmic corrections due to hydrodynamics, which increase the local
exponent on the order of $10\%$.  

\begin{figure}[!t]
\includegraphics*[scale=1]{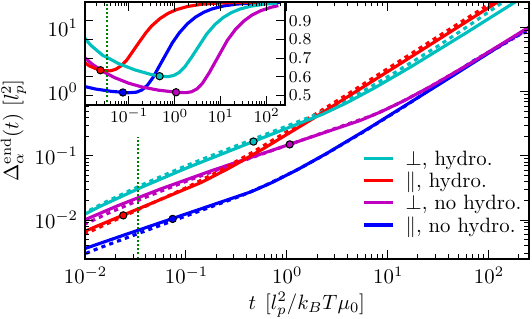}
\caption{End-point MSD $\Delta^\text{end}_\alpha(t)$,
  $\alpha=\perp,\parallel$, for a chain with $L=100a$, $L/l_p=5$,
  $F=20\:k_B T/l_p$.  The top two curves include hydrodynamic
  interactions, while the bottom two are free-draining.  MFT results
  are shown as solid lines, BD simulations as dashed lines. The inset
  shows the local slopes of the MFT curves in the log-log plot.  MFT
  relaxation times $\tau_{\alpha 1}$ are marked by circles, while the
  crossover time $t^\ast$ is marked by a vertical dashed
  line.}\label{hydro}
\end{figure}

Note that even in the strongly stretched limit, $F l_p / k_B T \gg 1$,
where the polymer is nearly straight, hydrodynamics is significant.
\ref{hydro} shows MFT and BD results for
$\Delta^\text{end}_\alpha$ with and without hydrodynamics for a chain
where $L=100a$, $L/l_p =5$, and $F = 20$ $k_BT/l_p$.  The MSD
components in the two cases cannot be related through a simple time
rescaling: for $t^\ast \ll t \lesssim \tau_{\alpha 1}$, we see clearly
the expected $t^{1/2}$ behavior for the free-draining chain (local
slopes are shown in the inset), while the exponent is pushed up to
$\approx 0.6-0.7$ with hydrodynamics.  While a careful WBA analysis
\cite{Granek1997} can include hydrodynamics and account for some of
the crossover effects, it is less quantitatively accurate than the MFT
for weaker forces and more flexible chains.  We will return to this
issue in Sec.~\ref{results:wbacomp}, where we make a direct comparison
of the two theories.

\begin{figure}[!t]
\centering\includegraphics*[scale=1]{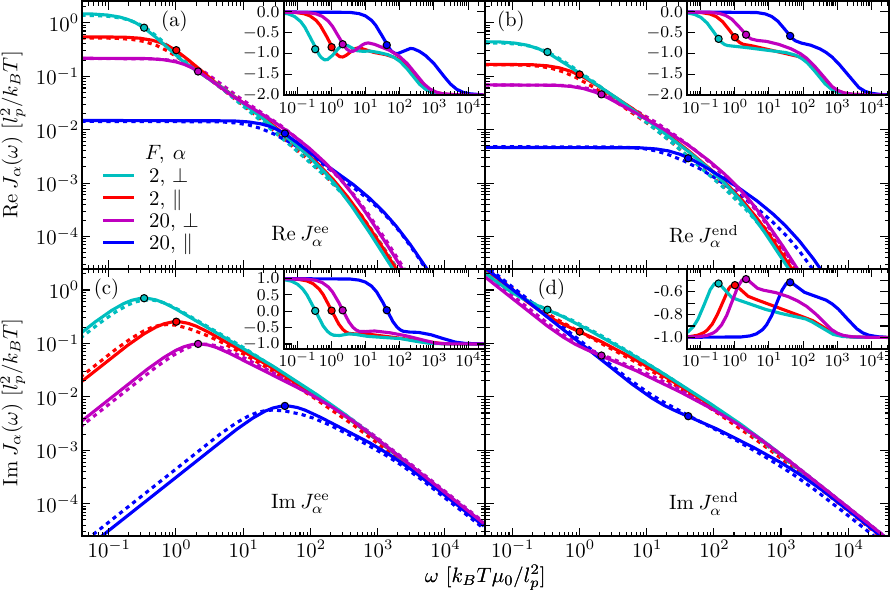}
\caption{Response functions for a polymer with $L=100a$, $L/l_p = 5$:
  (a) $\text{Re}\:J^\text{ee}_\alpha(\omega)$; (b)
  $\text{Re}\:J^\text{end}_\alpha(\omega)$; (c)
  $\text{Im}\:J^\text{ee}_\alpha(\omega)$; (d)
  $\text{Im}\:J^\text{end}_\alpha(\omega)$.  Solid lines are the
  anisotropic MFT results, while dashed lines are taken from BD
  simulations.  In all cases long-range hydrodynamic interactions are
  taken into account, and results are given for
  $\alpha=\perp,\parallel$ at two forces, $F=2$ and $20\: k_B T/l_p$.
  Filled circles mark the inverse relaxation times $\tau^{-1}_{\alpha 1}$, derived
  from the MFT, while the insets show the local slopes of the MFT
  curves in the log-log plots.}\label{alpfull}
\end{figure}

Using the fluctuation-dissipation theorem, the MSD can reveal the
viscoelastic properties of the chain: Fourier transforming the MSD
functions gives the imaginary parts of the end-to-end and self
response functions of the polymer end-points, 
\begin{equation}\label{bdc0}
\text{Im}\: J^\text{end}_\alpha(\omega) = - \frac{i \omega}{2 k_B T} \Delta^\text{end}_\alpha(\omega), \qquad \text{Im}\: J^\text{ee}_\alpha(\omega) = - \frac{i \omega}{2 k_B T} \Delta^\text{ee}_\alpha(\omega), 
\end{equation}
which are defined as:
\begin{equation}\label{bdc1}
J^{\text{end}}_\alpha(\omega) = \frac{\delta r_\alpha(L,\omega)}{f_\alpha(\omega)},\qquad J^{\text{ee}}_\alpha(\omega) =
\frac{\delta R_\alpha(\omega)}{f_\alpha(\omega)}.
\end{equation}
Here $\delta r_\alpha(L,\omega)$ and $\delta R_\alpha(\omega)$ are the
complex oscillation amplitudes resulting from a small force
$f_{\alpha}(\omega) = f_0 \exp(-i\omega t)$ applied to one end
($\mu=\text{end}$) or between both ends ($\mu=\text{ee}$) of the
chain, in addition to the prestretching tension $F$.  From the MFT
solution, Eqs.~\eqref{eq:d19}-\eqref{eq:d20}, one can express
$J^{\mu}_\alpha(\omega)$ as a sum over normal mode contributions,
\begin{equation}\label{bdc2}
J^{\mu}_\alpha(\omega) =
\delta_{\mu,\text{end}} \frac{i D_\alpha}{\omega k_B T} + \sum_{n=1}^{M-1}\frac{
A^{\mu}_{\alpha n}}{1-i \omega \tau_{\alpha n}},
\end{equation}
 with center-of-mass diffusion parameters $D_\alpha$, relaxation times
 $\tau_{\alpha n}$, and coefficients $A^\mu_{\alpha n}$.  The mode
 number cutoff $M = L/8a$ is chosen to roughly model the discrete
 nature of the chain at length scales comparable to the bead diameter,
 but results at larger length scales are independent of the
 cutoff~\cite{Hinczewski2009}.  

\ref{alpfull} shows the real and imaginary parts of
$J^\text{ee}_\alpha(\omega)$ and $J^\text{end}_\alpha(\omega)$, for
the same parameters as in \ref{msdalp}, compared to the results
extracted from BD simulations.  The good quantitative agreement with
BD in the time domain is carried over to frequency space: the
simulation trends are accurately reproduced by the MFT.  For $\omega
\ll \tau_{\alpha 1}^{-1}$, we see a mainly elastic end-to-end
response, $J^\text{ee}_\alpha \approx A^\text{ee}_{\alpha 1}(1+i
\omega \tau_{\alpha 1})$, with an effective spring constant $
(A^\text{ee}_{\alpha 1})^{-1}$.  The self response of the end-point at
these small frequencies is proportional to the center-of-mass
mobility, $J^\text{end}_\alpha \approx i D_\alpha/\omega k_B T$.  For
$\omega \gtrsim \tau_{\alpha 1}^{-1}$ we pass into the more
interesting high-frequency regime governed by the complex nature of
normal mode relaxation under tension and hydrodynamic interactions (up
to the ultraviolet cutoff at $\tau_{\alpha M}^{-1}$, above which the
discreteness of the chain dominates).  The effects of tension in this
regime have been directly observed in cytoskeletal networks through
microrheology~\cite{Caspi1998,Mizuno2007}: with increasing force the
dynamic compliance is reduced, and the high-frequency scaling changes
from $\omega^{-3/4}$ (the behavior of a relaxed semiflexible network)
to $\omega^{-1/2}$.  Qualitatively, we find both of these stiffening
effects in our MFT results in \ref{alpfull}: the magnitudes of
$J^\text{ee}_\alpha(\omega)$ and $J^\text{end}_\alpha(\omega)$
generally decrease with with force, and the $\omega$ scaling
(indicated by the local slopes) is shifted.  Unlike a network, where
hydrodynamics is screened, in the single polymer case the long-range
interactions modify the local slopes: rather than $-3/4$ and $-1/2$,
we see $\approx -0.8$ at weak force changing to $\approx -0.6$ at
strong force (most clearly evident in the insets to the Re and Im
$J^\text{end}_\alpha$ panels in the right column of \ref{alpfull}, in
the plateau-like slope region between $\tau_{\alpha 1}^{-1} \lesssim
\omega \lesssim \tau_{\alpha M}^{-1}$).  This correction, along with
the full crossover behavior of the imaginary response---proportional
to the power spectral density (PSD)---should be observable in future
nanorheology experiments for single semiflexible chains (i.e. the AFM
techniques already used to extract the PSD of flexible
polymers~\cite{Sakai2002,Khatri2007}, or optical tweezer methods).

\subsection{Comparison with experimental relaxation times of stretched DNA}\label{results:dnacomp}

\begin{figure}[!t]
\includegraphics*[scale=1]{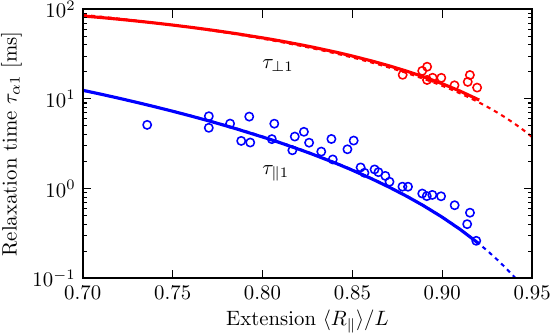}
\caption{The longest relaxation times parallel (blue) and
  perpendicular (red) to the force direction, $\tau_{\parallel 1}$ and
  $\tau_{\perp 1}$, for a single double-stranded DNA (contour length
  $L = 16.4$ $\mu$m) stretched in an optical tweezer.  Experimental
  data (open circles) from Ref.~\citenum{Meiners2000} is plotted as a
  function of relative longitudinal chain extension $\langle
  R_\parallel\rangle / L$.  The predictions of the anisotropic MFT,
  calculated without fitting parameters, are drawn as solid curves.
  The results of the simple scaling theory, Eq.~\eqref{eqh9}, described
  in Sec.~\ref{results:dnacomp} are plotted as dashed curves.  The
  constant prefactors in the scaling results are fitted to the MFT
  curves, yielding best-fit values of $c_\parallel = 0.122$ and
  $c_\perp =0.100$.  Consequently the MFT and scaling curves largely
  overlap.}\label{quakecomp}
\end{figure}

One dynamical quantity for which single-molecule experimental results
already exist is the largest relaxation time, $\tau_{\alpha 1}$.
Meiners and Quake have extracted the transverse and longitudinal
relaxation times, $\tau_{\perp 1}$ and $\tau_{\parallel 1}$, from
thermal fluctuations of a double-stranded DNA chain ($L = 16.4$
$\mu$m) stretched within an optical tweezer~\cite{Meiners2000}.  The
data is plotted as a function of longitudinal chain extension $\langle
R_\parallel \rangle/L$ in \ref{quakecomp} (open circles).  The
anisotropic MFT predictions are drawn as solid curves.  Again there
are no fitting parameters, since the theory depends only on the given
value of $L$, $l_p = 50$ nm, $a=1$ nm, $T = 298$ K, and $\eta = 0.891$
$\text{mPa}\cdot\text{s}$.  The agreement with experiment is very
good, with average deviations of 26\% for $\tau_{\parallel 1}$ and
18\% for $\tau_{\perp 1}$.

The behavior of the relaxation times in this example, for a chain that
is mostly extended, can be modeled through a simple scaling
theory~\cite{Hatfield1999,Meiners2000}.  If we treat the $n=1$ mode of
the polymer effectively as the oscillation of a spring, we can write
$\tau_{\alpha 1} = (\mu_{\alpha} k_\alpha)^{-1}$.  Here $\mu_\alpha$
and $k_\alpha$ is the effective mobility and spring constant
respectively.  When the chain is near maximum extension, the
mobilities can be estimated as those of a thin rod of diameter $d=2a$,
namely $\mu_\parallel = \ln(L/d)/(2\pi \eta L)$, $\mu_\perp =
\mu_\parallel/2$, to leading order.  To get the spring constants, the
starting point is the approximate Marko-Siggia interpolation
formula~\cite{Marko1995} relating the tension $F$ felt by a
semiflexible chain to its average end-to-end extension $R_\text{ee}$
along the $z$ axis:
\begin{equation}\label{eqh6}
F(R_\text{ee}) \approx \frac{k_BT}{l_p}\left[\frac{R_\text{ee}}{L} + \frac{1}{4(1-R_\text{ee}/L)^2} - \frac{1}{4}\right].
\end{equation}
The force magnitude $F$ is related to the polymer free energy ${\cal
  F}$ through $F = \partial {\cal F}/\partial R_\text{ee}$, and thus
the effective longitudinal spring constant $k_\parallel = \partial^2
{\cal F}/\partial R_\text{ee}^2 = \partial F/\partial R_\text{ee}$.
We can then estimate $k_\parallel$ from Eq.~\eqref{eqh6}:
\begin{equation}\label{eqh7}
\begin{split}
k_\parallel = \frac{k_BT}{l_p}\left[\frac{1}{L} + \frac{1}{2L(1-R_\text{ee}/L)^3}\right].
\end{split}
\end{equation}
For the transverse direction, we use the following
relation~\cite{Hatfield1999}: if a polymer stretched along $z$, with
extension $R_\text{ee}$, has one end displaced by a small transverse
distance $\delta R_\perp$, the restoring force $\delta F_\perp =
(\delta R_\perp/\sqrt{R_\text{ee}^2+\delta
  R_\perp^2})F(\sqrt{R_\text{ee}^2+\delta R_\perp^2})$.  To leading order in
$\delta R_\perp$, this gives $\delta F_\perp = \delta R_\perp
F(R_\text{ee})/R_\text{ee}$, or equivalently $k_\perp =
F(R_\text{ee})/R_\text{ee}$.  From Eq.~\eqref{eqh6} we have:
\begin{equation}\label{eqh8}
\begin{split}
k_\perp = \frac{k_BT}{l_p}\left[\frac{1}{L} + \frac{1}{4R_\text{ee}(1-R_\text{ee}/L)^2} - \frac{1}{4R_\text{ee}}\right].
\end{split}
\end{equation}
Putting everything together we get the following expressions for the
relaxation times in terms of $R_\text{ee}$:
\begin{equation}\label{eqh9}
\begin{split}
\tau_{\parallel 1} &= c_\parallel \frac{2\pi \eta L l_p}{k_BT \ln(L/d)}\left[\frac{1}{L} + \frac{1}{2L(1-R_\text{ee}/L)^3}\right]^{-1},\\
\tau_{\perp 1} &= c_\perp \frac{4\pi \eta L l_p}{k_BT\ln(L/d)}\left[\frac{1}{L} + \frac{1}{4R_\text{ee}(1-R_\text{ee}/L)^2} - \frac{1}{4R_\text{ee}}\right]^{-1}.
\end{split}
\end{equation}
Since this is a scaling argument, we expect the results to be
approximately valid up to some constant prefactors, which we denote
$c_\parallel$ and $c_\perp$.  In fact, Eq.~\eqref{eqh9}, plotted as
dashed lines in \ref{quakecomp}, can be made to overlap the MFT curves
almost perfectly, with best-fit prefactors of $c_\parallel = 0.122$
and $c_\perp=0.100$.  These are very close to the prefactor $1/\pi^2
\approx 0.101$ estimated in Ref.~\citenum{Meiners2000} from the
fluctuation-dissipation theorem.  The relaxation times in
Eq.~\eqref{eqh9} can alternatively be expressed as scaling functions
of $F$,
\begin{equation}\label{eqh10}
\tau_{\parallel 1} = c_\parallel \frac{\pi \eta L^2 l_p}{2k_BT \ln(L/d)}\left(\frac{l_p F}{k_B T}\right)^{-3/2}, \quad \tau_{\perp 1} = c_\perp \frac{4\pi \eta L^2 l_p}{k_B T\ln(L/d)}\left(\frac{l_p F}{k_B T}\right)^{-1},
\end{equation}
valid in the large $F$ limit, $F l_p/k_B T \gg 1$.

As a side note, we have to be careful to assess the importance of
self-avoidance in cases where the chain contour $L \gg l_p$, since
this is neglected both in the MFT and the scaling argument.  If we
were to go to the limit of small forces, $F l_p / k_B T \ll 1$, and
extremely long chains, $L \gg l_p^3/a^2$, one expects to see the
influence of self-avoidance~\cite{NetzAndelman2003}.  In this
particular experimental example, neither of these conditions holds,
since $F l_p / k_B T \approx 3 - 50$ for the measured extensions, and
$L = 16.4\:\mu\text{m} \ll l_p^3/a^2 = 125\:\mu\text{m}$.

\subsection{Comparison with the weakly bending approximation}\label{results:wbacomp}

Finally, it is instructive to compare the anisotropic MFT results to
the WBA, to illustrate the range of applicability and relative
strengths of both approaches.  The traditional weakly bending
Hamiltonian for the transverse fluctuations of a chain with constant
backbone tension $F$ and persistence length $l_p$ is given
by~\cite{Granek1997,Hallatschek2005}:
\begin{equation}\label{wba1}
U^\perp_\text{WBA}=\frac{l_p k_BT}{2} \int ds\,\left(\partial_s \mb{u}_\perp(s)\right)^2 +\frac{F}{2} \int
ds\,\mb{u}_\perp^2(s).
\end{equation}
Comparing to Eq.~\eqref{amft1}, we see that $U^\perp_\text{WBA}$ is a
special case of the $\alpha = \perp$ component of the anisotropic
Gaussian Hamiltonian $U_\text{MFA}$, with $\epsilon_\perp = l_p k_B
T$, $\nu_\perp = F/2$, and $\nu_{0\perp} = 0$.  By making these
substitutions in the derivation of Sec.~\ref{amft:dyn} (confining
ourselves to the $\alpha=\perp$ part), and using the pre-averaged
hydrodynamic tensor $\mu^\perp_\text{avg}(l) = 2a\mu_0 \delta(l) +
3\Theta(l-2a)\mu_0/4 l$ (a special case of Eq.~\eqref{d6},
appropriate for a nearly rigid rod) we can recover the basic WBA
dynamical theory for chains under tension (i.e. Sec. 4.2 of Granek's
study~\cite{Granek1997}).  The main difference from
Ref.~\citenum{Granek1997} is that our normal modes incorporate the
correct boundary conditions, rather than being based on a Fourier
expansion which is strictly valid only far from the chain ends.  In
fact, at $F=0$ the normal modes derived in our way reduce to the
expected Arag\'on and Pecora expressions~\cite{Aragon1985}.  The
resulting dynamical equations based on $U^\perp_\text{WBA}$ yield
transverse observables like $\Delta_\perp^\text{ee}(t)$.  The
corresponding longitudinal quantities like
$\Delta_\parallel^\text{ee}(t)$ are derived in the WBA approach using
the approximate relation $u_\parallel(s,t)\approx 1 -
\mb{u}_\perp^2(s,t)/2$, valid when $\mb{u}_\perp^2(s,t)$ is small.

\begin{figure}[!t]
\centering\includegraphics*[scale=1]{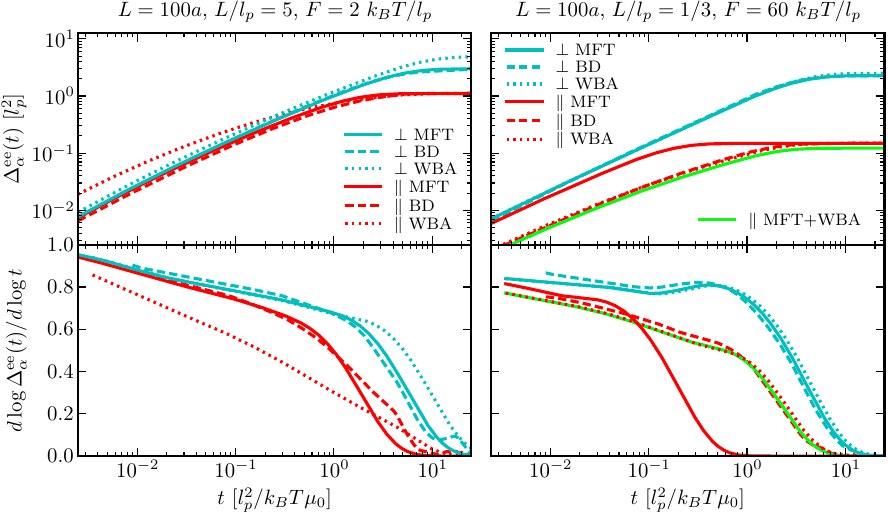}
\caption{Top: End-to-end MSD $\Delta^\text{ee}_\alpha(t)$,
  $\alpha=\perp$ (cyan), $\parallel$ (red); bottom: the corresponding
  local slope $d\log \Delta^\text{ee}_\alpha(t)/d\log t$.  Two
  different sets of chain parameters are shown in the two columns:
  $L=100a$, $L/l_p=5$, $F=2\:k_B T/l_p$ (left), and $L=100a$,
  $L/l_p=1/3$, $F=60\:k_B T/l_p$ (right).  Anisotropic MFT results are
  drawn as solid lines, BD simulations (with long-range hydrodynamic
  coupling) as dashed lines, and the WBA results (described in
  Sec.~\ref{results:wbacomp}) as dotted lines.  For the nearly rigid
  rod case shown on the right, the additional green curve marked
  MFT+WBA is an estimate for the longitudinal dynamics based on
  applying a WBA-like expansion to the transverse MFT, as described in
  App.~A.}\label{wbacomp}
\end{figure}

The WBA dynamical theory for the end-to-end MSD functions
$\Delta_\alpha^\text{ee}(t)$, $\alpha = \perp,\parallel$ is contrasted
to the anisotropic MFT and Brownian dynamics (BD) results in the left
panels of \ref{wbacomp} for a chain with $L = 100a$, $L/l_p = 5$, and
$F = 2$ $k_B T/l_p$.  Though the chain is stretched out for these
parameters, with small transverse fluctuations ($\langle \delta
R_\perp^2 \rangle/2L^2 \approx 0.06$), the WBA performs worse than the
MFT when compared to the simulation results.  Average deviations
between the WBA and BD in the time range shown are generally 5-10
times larger than the analogous deviations between the MFT and BD,
both for transverse and longitudinal components.  In the case of the
longitudinal WBA, the deviations from the simulations at very short
times may partially be accounted for by an effect which is not present
in our formulation: we do not include corrections for longitudinal
friction~\cite{Everaers1999}, which are expected to be relevant for
times $t \ll k_B T l_p /\mu_0 F^2 = 0.25\: l_p^2 /k_B
T\mu_0$, and which are present in more sophisticated implementations
of the
WBA~\cite{Hallatschek2005,Obermayer2007,Hiraiwa2008,Obermayer2009}
(though these more advanced approaches do not include long-range
hydrodynamic interactions, which we incorporate into our version of
the WBA).  At the very largest times plotted, small oscillations in
the slopes calculated from BD are artifacts due to insufficiently
converged simulation data.  Since the slopes are numerical derivatives
of the MSD functions, they are particularly sensitive to noise.
However this issue does not affect the clear deviations in slopes for
$t \lesssim 1$ $l_p^2/k_B T\mu_0$.

The WBA becomes highly accurate in the limit of extremely large
force or large persistence length, when the chain is almost fully
extended.  We show this in the right panel of \ref{wbacomp}, for
parameters $L = 100a$, $L/l_p = 1/3$, $F = 60$ $k_B T/l_p$, where the
WBA and BD results now nearly overlap.  However, here we see a
limitation of the anistropic MFT: for a system that is nearly a rigid
rod, no Gaussian model will be able to capture the longitudinal
dynamics.  The MFT underestimates $\tau_{\parallel 1}$ by an order of
magnitude, with $\Delta_\parallel^\text{ee}(t)$ saturating to
equilibrium much quicker than the BD result.

On the other hand, the transverse MFT is still remarkably precise, deviating $<
7\%$ from the BD curve throughout the entire time range.  This is not
surprising, since the coefficients in the transverse MFT, which are
dependent on the parameters of the chain, behave like $\epsilon_\perp
\to l_p k_B T$, $\nu_\perp \to F/2$ as $L/l_p \to 0$ and/or $F \to
\infty$ (the $\nu_{0\perp}$ term has a negligible effect in these
limits).  These trends are in line with the results in \ref{fig1} for large
$F$.  In other words, the transverse MFT converges to the WBA
Hamiltonian in the stiff rod limit, and in this sense the transverse
MFT is a general theory that contains the WBA as a limiting case.

Knowing the breakdown of the longitudinal MFT in the asymptotic
limit, we can actually incorporate a fix: using the transverse MFT
results in combination with the WBA to estimate longitudinal
quantities (taking advantage of the fact that the transverse MFT works
well in all regimes).  The key relation is $u_\parallel(s,t) \approx 1
- \mb{u}_\perp^2(s,t)/2$, valid in the stiff limit.  Using the
transverse MFT estimate of $\mb{u}_\perp^2(s,t)$, one can derive a
first-order perturbation expansion for $\Delta^\text{ee}_\parallel(t)$
(details are in App.~A), yielding the green curve in \ref{wbacomp}.
This gives a much better agreement with BD ($<25\%$ deviation) than
the original MFT.  As described above, this fix for the longitudinal
theory is only necessary for the rigid rod limit; otherwise the
original MFT is the preferred choice.

\section{Conclusions}

In summary, we have developed an anisotropic MFT for the dynamics of
semiflexible chains under tension, whose most notable feature is
quantitative accuracy over a broad range of dynamical
regimes---verified through BD simulations and comparison to
single-molecule measurements on DNA.  The theory precisely captures
the interplay of backbone rigidity, long-range hydrodynamic
interactions, and large-scale motion of the polymer contour that
contribute to the challenge of modeling semiflexible polymer dynamics.

Understanding kinetics of single stretched chains is interesting in
itself (or as the first step toward more elaborate theories of
stressed networks), but it can also be exploited in other contexts:
optical tweezer force-clamp experiments depend sensitively on the
dynamical response of the DNA handles that are attached to the object
of interest, whether a nucleic acid hairpin or
protein~\cite{Hyeon2008}.  A prerequisite for filtering out the handle
effects, in order to extract the intrinsic properties of the
biomolecule in the clamp, is an accurate theory for the handle
dynamics.

The simple Gaussian form of the anisotropic MFT has its own
advantages: it allows easy analytical computation of various
additional quantities like Green's functions describing the stochastic
time evolution of the polymer.  For the $F=0$ case, this fact has
already been exploited to model diffusion-limited reactions between a
DNA-binding protein and its target site on the DNA, using the MFT to
incorporate contour fluctuations and hydrodynamic
effects~\cite{vonHansen2010}.  For $F\ne 0$, the Green's function
formalism will allow precise estimates in reaction-diffusion systems
involving semiflexible components under tension, like motor proteins
stepping under load---one of many macromolecular systems where our
approach can be fruitfully applied.

\begin{acknowledgments}
The authors thank the Feza G\"ursey Institute for use of the Gilgamesh computing cluster.
\end{acknowledgments}

\section*{Appendix A:  WBA estimate for $\Delta_\parallel^\text{ee}(t)$ based on the transverse MFT results}

For small deviations from the rigid rod limit, the transverse and
longitudinal tangent vectors of the WLC can be related as:
$u_\parallel(s,t) = \sqrt{1-\mb{u}_\perp^2(s,t)} \approx 1 -
\mb{u}_\perp^2(s,t)/2 + \cdots$, where $\mb{u}_\perp = (u_x,u_y)$.
This is the fundamental equation for the WBA, and it allows one to
derive certain aspects of the longitudinal dynamics assuming the
transverse dynamics are known, specifically the behavior of
$\mb{u}_\perp^2(s,t)$.  As seen in Sec.~\ref{results:wbacomp}, the
anisotropic MFT provides a highly accurate prediction for the
transverse end-to-end MSD even for very stiff chains, so we can
exploit the reliability of the transverse dynamical theory through the
WBA approach.

We focus on finding an estimate for longitudinal end-to-end MSD
$\Delta_\parallel^\text{ee}(t)$, though the method is generalizable to
other dynamical quantities.  $\Delta_\parallel^\text{ee}(t)$ can be
expressed as $\Delta_\parallel^\text{ee}(t) = 2(C_\parallel(0) -
C_\parallel(t))$, where the correlation function $C_\parallel(t)$ is
given by:
\begin{equation}\label{wb1}
C_\parallel(t) = \langle R_\parallel(t) R_\parallel(0) \rangle - \langle R_\parallel \rangle^2 = \int_0^L ds \int_0^L ds^\prime \langle u_\parallel(s,t)u_\parallel(s^\prime,0) \rangle - \left[\int_0^L ds\, \langle u_\parallel(s,0)\rangle \right]^2.
\end{equation}
Here we have used the fact that $R_\parallel(t) = \int_0^L
ds\,u_\parallel(s,t)$.  Plugging in the first-order expansion
$u_\parallel(s,t)\approx 1 - \mb{u}_\perp^2(s,t)/2$, we get an
expression for $C_\parallel(t)$ involving averages over various
products of $\mb{u}_\perp^2(s,t)$.  From the normal mode expansion in
Sec.~\ref{amft:dyn} we know that $\mb{u}_\perp (s,t) = \partial_s
\mb{r}_\perp(s,t) = \sum_n \mb{P}_{\perp n}(t)
\Psi^{\perp\prime}_{n}(s)$, where $\mb{P}_{\perp n} = (P_{xn},P_{yn})$
and $\Psi^{\perp\prime}_{n}(s) \equiv \partial_s \Psi^\perp_{n}(s)$.
Thus all averages over $\mb{u}_\perp^2(s,t)$ are averages over the
normal mode amplitudes $\mb{P}_{\perp n}(t)$, and these can be
directly calculated from Wick's theorem and the solution of
Eq.~\eqref{d17} for the $\perp$ components.  The final result for
$C_\parallel(t)$ at order ${\cal O}(\mb{u}_\perp^2)$ has the form:
\begin{equation}\label{wb2}
\begin{split}
C_\parallel(t) =& \sum_{k,l} f_k(t) f_l(t) M_{kl}^2,
\end{split}
\end{equation}
where:
\begin{equation}\label{wb3}
\begin{split}
f_k(t)&= \frac{k_B T \Theta_{\perp k}}{\Lambda_{\perp k}} \exp(-\Lambda_{\perp k} t),\qquad M_{kl}=\int_0^L ds\,\Psi^{\perp\prime}_k(s) \Psi^{\perp\prime}_l(s).
\end{split}
\end{equation}
Thus $C_\parallel(t)$ can be expressed entirely in terms of quantities
from the $\perp$ MFT solution: the parameters $\{\Lambda_{\perp n},
\Theta_{\perp n}\}$ and the normal modes $\{ \Psi^{\perp}_n(s) \}$.
Numerical evaluation of Eqs.~\eqref{wb2}-\eqref{wb3} yields
$C_\parallel(t)$ and hence $\Delta^\text{ee}_\parallel(t)$.

\bibliography{polyref}

\end{document}